%% file: main.tex
\documentclass[a4paper,11pt,dvipsnames]{article}

\pdfoutput=1 

\usepackage{jheppub} 

\usepackage{shuffle}
\usepackage{comment}
\usepackage[T1]{fontenc} 
\usepackage{tikz}
\usepackage{tikz-3dplot}
\usepackage{xcolor}
\usetikzlibrary{patterns, decorations.markings, arrows}
\usetikzlibrary{arrows.meta}
\usepackage{mathabx}
\usepackage[export]{adjustbox}
\allowdisplaybreaks
\usepackage[backgroundcolor=green!5, linecolor=blue!60]{todonotes}

\newcommand{\la}{\langle}
\newcommand{\ra}{\rangle}
\newcommand{\ang}[1]{\langle #1\rangle}

\title{Non-planar BCFW Grassmannian Geometries}

\author{Shruti Paranjape,}\emailAdd{sparanjape@ucdavis.edu}
\author{Jaroslav Trnka,}\emailAdd{trnka@ucdavis.edu}
\author{Minshan Zheng}\emailAdd{mszheng@ucdavis.edu}

\affiliation{Center for Quantum Mathematics and Physics (QMAP),\\Department of Physics, University of California, Davis, CA 95616, USA}

\abstract{In this paper, we study non-adjacent BCFW recursion relations and their connection to positive geometry. For an adjacent BCFW shift, the $n$-point N$^k$MHV tree-level amplitude in 
${\cal N}=4$ SYM theory is expressed as a sum over planar on-shell diagrams, corresponding to canonical “dlog” forms on the cells in the positive Grassmannian $G_+(k,n)$. Non-adjacent BCFW shifts naturally lead to an expansion of the amplitude in terms of a different set of objects, which do not manifest the cyclic ordering and the hidden Yangian symmetry of the amplitude. We show that these terms can be interpreted as dlog forms on the {\it non-planar Grassmannian geometries}, generalizing the cells of the positive Grassmannian $G_+(k,n)$ to a larger class of objects which live in $G(k,n)$. We focus mainly on the case of NMHV amplitudes and discuss in detail the Grassmannian geometries. We also propose an alternative way to calculate the associated on-shell functions and dlog forms using an intriguing connection between Grassmannian configurations and the geometry in the kinematical space.}

\begin{document}

\maketitle

\section{Introduction}

Recent years have seen enormous progress in our understanding of scattering amplitudes in the planar ${\cal N} = 4$ super Yang-Mills (SYM) theory. These discoveries include the powerful methods of recursion relations for tree-level amplitudes and loop integrands \cite{Britto:2004ap,Britto:2005fq,Arkani-Hamed:2010zjl}, multi-loop calculations using unitarity methods \cite{Bern:1994zx,Bern:1994cg,Bern:2005iz,Bern:2009kd,Bern:2008ap,Bern:2012uc,Bern:2018jmv,Carrasco:2021otn,Bourjaily:2016evz,Bourjaily:2017wjl}, the discovery of hidden dual conformal and Yangian symmetries \cite{Drummond:2006rz,Drummond:2008vq,Drummond:2009fd}, the Wilson loop/amplitude correspondence \cite{Alday:2007hr,Drummond:2007cf,Caron-Huot:2010ryg}, the flux-tube approach for amplitudes at arbitrary coupling \cite{Basso:2013vsa,Basso:2014hfa,Basso:2015uxa}, the hexagon bootstrap \cite{Dixon:2011nj,Caron-Huot:2019vjl,Caron-Huot:2019bsq,Basso:2020xts,Dixon:2022rse,Dixon:2021tdw}, connections to cluster algebras \cite{Goncharov:2010jf,Golden:2013xva,Drummond:2014ffa,Drummond:2017ssj,Drummond:2019qjk,Mago:2020kmp,Ren:2021ztg}, momentum twistor methods \cite{Hodges:2009hk,Hodges:2010kq,Arkani-Hamed:2010wgm,Arkani-Hamed:2010pyv,Drummond:2010mb,Bourjaily:2013mma,Arkani-Hamed:2014dca,Bourjaily:2015jna}, and the new geometric formulation using positive Grassmannian \cite{Arkani-Hamed:2009ljj,Arkani-Hamed:2009nll,Arkani-Hamed:2009kmp,Arkani-Hamed:2009pfk,Mason:2009qx,Arkani-Hamed:2012zlh}, on-shell diagrams \cite{Arkani-Hamed:2012zlh,Franco:2013nwa,Arkani-Hamed:2014bca,Franco:2015rma,Bourjaily:2016mnp,Heslop:2016plj,Herrmann:2016qea,Farrow:2017eol,Armstrong:2020ljm,Cachazo:2018wvl} and the Amplituhedron \cite{Arkani-Hamed:2013jha,Arkani-Hamed:2013kca,Arkani-Hamed:2017vfh,Damgaard:2019ztj,Ferro:2020ygk,Herrmann:2022nkh,Herrmann:2022nkh}. The Amplituhedron picture has been successfully used to obtain various explicit results including all-loop order quantities \cite{Bai:2014cna,Franco:2014csa,Ferro:2016zmx,Ferro:2015grk,Ferro:2016ptt,Ferro:2018vpf,Ferro:2020lgp,Herrmann:2020oud,Herrmann:2020qlt,Kojima:2020tjf,Arkani-Hamed:2021iya,Rao:2019wyi,Kojima:2020gxs,Dian:2021idl,Dian:2022tpf}, and has also been an active area of research in pure mathematics \cite{Lam:2014jda,Karp:2016uax,Lukowski:2020dpn,Parisi:2021oql,Williams:2021zph,Moerman:2021cjg,Even-Zohar:2021sec}.

The key question is whether some of these developments can be generalized beyond the planar sector of the ${\cal N}=4$ SYM perturbative S-matrix. Naively, planarity is essential, as it implies a cyclic ordering of the amplitude without which we can not talk about dual conformal symmetry, integrability methods, the positive Grassmannian, or the Amplituhedron. Nevertheless, we have seen that many of the properties indeed survive beyond the planar limit, evidencing that the new ideas extend to the full ${\cal N}=4$ SYM S-matrix. For instance, the absence of poles at infinite momenta in the multi-loop non-planar integrands indicates that the analogue of dual conformal symmetry for full ${\cal N}=4$ SYM should exist \cite{Arkani-Hamed:2014via,Bern:2014kca,Bern:2015ple,Bern:2017gdk,Bern:2018oao,Bourjaily:2018omh,Chicherin:2018wes,Bourjaily:2019gqu}. The study of non-planar amplitudes in ${\cal N}=4$ SYM also pertains to a larger question of the role of planarity in the study of scattering amplitudes. Understanding how to do without planar variables and cyclic symmetry is crucial for addressing open questions in perturbative gravity where the graviton amplitudes are intrinsically non-planar. And indeed there are many intriguing properties of graviton amplitudes such as color-kinematics duality \cite{Bern:2008qj,Bern:2010ue,Bern:2019prr}, surprisingly compact formulas for tree-level amplitudes \cite{Bern:1998sv,Elvang:2007sg,Drummond:2009ge,Mason:2009afn,Nguyen:2009jk,Hodges:2011wm,Hodges:2012ym,Trnka:2020dxl} or enhanced UV behavior of loop amplitudes \cite{Bern:2012gh,Bern:2014sna,Bern:2017lpv,Herrmann:2018dja,Edison:2019ovj}, all of which suggest new formulations for amplitudes that extend beyond the planar sector. 

One interesting class of non-planar objects studied extensively in the past are on-shell diagrams. These gauge invariant functions appear as discontinuities of loop amplitudes across branch-cuts, where all intermediate particles are taken to be on-shell. There exists a fundamental connection between on-shell diagrams in quantum field theory and Grassmannian geometry. For an $n$-point N$^{k-2}$MHV amplitude, each planar on-shell diagram corresponds to a cell of the positive Grassmannian $G_+(k,n)$, whereas a generic non-planar diagram is associated with a general cell in $G(k,n)$ without any positivity properties. The non-planar on-shell diagrams were partially classified in a series of papers \cite{Arkani-Hamed:2014bca,Bourjaily:2016mnp}. But much remains to be understood about non-planar on-shell diagrams. On-shell diagrams can be computed as products of tree-level amplitudes from the definition. The same expressions are also reproduced as canonical differential forms associated with the corresponding cells in $G(k,n)$. In ${\cal N}=4$ SYM, the canonical forms are simple “dlog”-forms in both the planar and non-planar cases. However, while the expressions for planar on-shell diagrams are invariant under dual super-conformal (and also Yangian) symmetry, the non-planar on-shell diagrams have no such symmetry a priori. It natural to ask if, and under what conditions, any part of the dual conformal symmetry survives in the non-planar sector. 

A hopeful starting point for investigating the problem of residual symmetry is the special class of non-planar on-shell diagrams which serve as the building blocks in the non-adjacent BCFW representation of scattering amplitudes. The tree-level BCFW recursion relations for cyclically ordered amplitudes can be realized diagrammatically in terms of on-shell diagrams, where each term in the recursion corresponds to one (or multiple) on-shell diagrams. These on-shell diagrams are not generic, as they originate from gluing together simpler diagrams that represent the two subamplitudes which are separated by an additional “BCFW bridge”. A very simple application of the recursion relations to gluons allows us to write very compact formulae for ${\cal N}=4$ SYM tree-level amplitudes. In the usual implementation of BCFW recursions, we use exclusively adjacent shifts, i.e.  BCFW bridges attached to two adjacent legs, leading to highly efficient expansions in \textit{planar} on-shell diagrams with manifest Yangian symmetry. 
But nothing prevents the choice of non-adjacent legs for BCFW shifts; in this case, the recursion relations give rise to expansions in terms of generally non-planar on-shell diagrams. 

In this paper we initiate a systematic study of the connection between non-planar on-shell diagrams arising in the context of non-adjacent BCFW shifts and the Grassmannian geometry. We identify the \emph{BCFW cells} in the Grassmannian $G(k,n)$ associated with these diagrams by looking at configurations of $n$ points in $\mathbb{P}^{k{-}1}$ as encoded in the representative $C$-matrices. While not retaining the cyclicity and convexity of their planar counterparts, the configurations underlying the non-planar on-shell diagrams generated by non-adjacent BCFW shifts turn out to form a surprisingly simple subset. We find a new way to extract on-shell functions directly from the Grassmannian configurations, bypassing the standard procedures of computation. The key insight, drawn from analyses of the geometry of MHV amplitudes and Parke-Taylor factors in the kinematic space, is to make connection between constrained Grassmannian configurations and singularities in external kinematic data. Exploiting this correspondence, we present a closed-form formula for the non-adjacent BCFW representation of tree-level amplitudes in terms of certain non-planar on-shell functions, analogous to the familiar planar R-invariants. Using the Kleiss-Kuijf (KK) relations, we show that the non-planar on-shell functions can be expressed as linear combinations of their planar counterparts, which is reminiscent of a similar statement for the MHV on-shell diagrams \cite{Arkani-Hamed:2014bca}. We mainly focus on NMHV amplitudes, but also show some N$^2$MHV examples and outline the generalization for any $k$.

The paper is organized as follows:  In Section \ref{sec:background}, we review the connection between tree-level BCFW recursion and cells in the positive Grassmannian. In Section \ref{sec:holoform}, we propose a new holomorphic formula for planar BCFW on-shell functions, extracted directly from the Grassmannian geometry as viewed in the kinematical space. In Section \ref{sec:nonadjBCFW}, we identify the non-planar Grassmannian geometries associated with non-adjacent BCFW terms in the NMHV amplitudes. In Section \ref{sec:nonplanarR}, we obtain the on-shell functions for these non-planar geometries which extend the formula in Section \ref{sec:holoform} to include new types of objects. We show that these can nevertheless be re-expressed as a linear combination of planar $R$-invariants with different orderings. In Section \ref{sec:dlog}, we further exploit the geometric picture and find a holomorphic dlog-form representation of the on-shell functions  directly in the kinematical space. In Section \ref{sec:higherk}, we extend the NMHV discussion in Sections \ref{sec:holoform}-\ref{sec:dlog} to the N$^2$MHV amplitudes and beyond. We end with the conclusions and outlook in Section \ref{sec:discuss}.

\section{Background: From BCFW to Positive Geometry}
\label{sec:background}

We consider the $n$-point N$^k$MHV gluon scattering amplitudes in ${\cal N}=4$ SYM theory. (Note that at tree-level, gluon amplitudes are identical for any number of supersymmetries ${\cal N}$, so here the maximal supersymmetry is introduced mainly for book-keeping reasons.) At tree-level, we can decompose the amplitude as a sum of cyclically ordered amplitudes,
\begin{equation}
{\cal A}_{n}^{(k)} = \sum_\pi {\rm Tr}(T^{a_1}T^{a_2}\dots T^{a_n})A_{n}^{(k)}(1,2,\dots n)\,,
\end{equation}
where the sum is over all permutations modulo cyclic ones. We focus on the amplitude $A_{n}^{(k)}\equiv A_{n}^{(k)}(1,2,\dots n)$ with a canonical ordering, which can be calculated using Britto-Cachaz-Feng-Witten (BCFW) recursion relations. We perform a BCFW shift,
\begin{equation}
    \lambda_{\hat i} = \lambda_i + z \lambda_j,\qquad \widetilde{\lambda}_{\hat j} = \widetilde{\lambda}_j - z \widetilde{\lambda}_i\,, \label{eq:shift}
\end{equation}
under which the amplitude $A_n^{(k)}(z)$ develops a dependence on the shift parameter. Since it scales like $A_n^{(k)}(z)={\cal O}(1/z)$ for $z\rightarrow\infty$, we can use the Cauchy formula to reconstruct the original amplitude $A_n^{(k)}=A_n^{(k)}(z=0)$, 
\begin{equation}
\label{cauchyformula}
    \oint \frac{A_n^{(k)}(z)\,dz}{z} = 0 \,\,\Rightarrow A_n^{(k)} = -\sum_{j} {\rm Res}_{z=z_j}\left[\frac{A_n^{(k)}(z)}{z}\right]\,.
\end{equation}
Based on tree-level unitarity, all poles of $A_n^{(k)}(z)$ correspond to $P_k(z)^2=0$ where $P_k$ is a (shifted) sum of adjacent external momenta, and the residues are products of lower-point amplitudes with shifted momenta,
\begin{equation}
    A_n^{(k)} = \sum_P \frac{A_{n_L}^{(k_L)}A_{n_R}^{(k_R)}}{P^2}\,, \label{BCFW}
\end{equation}
where $n_L+n_R=n-2$, $k_L+k_R=k-1$, and we sum over all factorization channels where particle $i$ is on the left side and particle $j$ on the right side. For \emph{adjacent} BCFW shifts, namely $(n1)$ shift,
\begin{equation}
\label{eq:BCFWrecursion}
    \lambda_{\hat n} = \lambda_n + z \lambda_1,\qquad \widetilde{\lambda}_{\hat 1} = \widetilde{\lambda}_1 - z \widetilde{\lambda}_n\,,
\end{equation}
the expression (\ref{BCFW}) simplifies to a sum over the single index $j$,
\begin{equation}
    A_n^{(k)} = \sum_j \frac{A_{n_L}^{(k_L)}(\hat{1},2,\dots j,I)A_{n_R}^{(k_R)}(I,j{+}1,\dots,\hat{n})}{P^2}\,,
\end{equation}
where the on-shell momentum $P$ of the internal line is given by
\begin{equation}
    P_I = \frac{(P)|n]\,\la 1|(P)}{\la 1|P|n]}\,,\quad \mbox{where}\,\, P = p_1+\dots+p_j.
\end{equation}
The adjacent BCFW recursion relations provide highly efficient representations of the tree-level amplitude involving a minimal number of terms. Furthermore, they preserve the cyclic ordering of external legs required for introducing momentum twistor variables \cite{Hodges:2009hk}. In the case of ${\cal N}=4$ SYM the adjacent BCFW expansions are then term-wise Yangian invariant, making this hidden symmetry of tree-level amplitudes manifest.

\subsection{From BCFW to On-shell Diagrams}

BCFW recursion relations have a natural diagrammatic realization in terms of \emph{on-shell diagrams}. An on-shell diagram is built from 3-point massless amplitudes connected with on-shell edges. There are two types of 3-particle massless amplitudes, represented by black and white black and white 3-pt vertices in the graph, which corresponds to MHV and $\overline{\text{MHV}}$ helicity conﬁguration, resp. In ${\cal N}=4$ SYM theory, all the various 3-pt amplitudes are packaged into two elementary superfunctions\footnote{While we are primarily interested in ${\cal N}=4$ SYM here, let us mention in the passing that on-shell diagrams can be defined for any quantum field theory. If we work in another theory, we just replace these expressions by other 3-point amplitudes. For theories with less than maximal supersymmetry, we need to add arrows on all legs indicating the helicities} 
\input{figures/ThreePoint}
Gluing these elementary building blocks together generate more complicated objects that we call on-shell diagrams. 

Functions corresponding to such graphs are called on-shell functions; they have the physical interpretation as residues of multi-loop integrands that computes discontinuities of loop amplitudes across branch-cuts, where all intermediate particles taken to be on-shell. If the
\emph{cuts} conditions localize all the loop variables, and impose no further constraints on external kinematics, e.g.
%
%
\input{figures/Cut}
we call the on-shell function (diagram) a \emph{leading singularity}. 
In this case $L=4 n_I$ where $L$ is the number of loops and $n_I$ number of internal propagators in the diagram; in other words, all $4L$ degrees of freedom in $L$ off-shell loop momenta $\ell_i$ are fixed by cut conditions. Note that the cuts conditions admit two sets of solutions for loop momenta $\ell_i$ and this information is also encoded in the diagram through collinearities of $\lambda$, $\widetilde{\lambda}$ in each vertex. 

To compute the on-shell function, we take the product of all three-point amplitudes in the diagram as evaluated on the cut solution 
and dress it with the Jacobian factor $J$ resulting from the elimination of internal variables. 


%
%
\begin{equation}
   \input{./figures/Onshell1}  \label{onshelldiag}
\end{equation}

Having introduced the on-shell diagrams, let us quickly review how to use them to implement BCFW recursion. The BCFW shift \eqref{eq:shift} of any on-shell object is furnished by a \emph{BCFW bridge}. If we denote the $n$-point amplitude $A_n$ by a grey blob schematically, then the $(n1)$-shifted amplitude $A_n(z)$ corresponds to attaching the following BCFW bridge:
%
%
\input{figures/BCFWshift}
Taking a residue of $A_n(z)$ corresponds to removing one internal edge from the diagram on the right. The Cauchy formula \eqref{cauchyformula} is then the statement that all the diagrams with one edge by removed sum to zero (see \cite{Arkani-Hamed:2012zlh} for details). 
%
%
\input{figures/GRT}
Thus we can then express the unshifted amplitude $A_n$ which corresponds to removing the edge in the BCFW bridge) as a sum of all the other on-shell diagrams, which are given by gluing together lower-point amplitudes (grey blobs) with the addition of the same BCFW bridge $(n1)$. 
Note that each grey blob in itself represents a sum of on-shell diagrams which are obtained recursively. 

Let us look at a few examples. At 4-point and 5-point, the MHV amplitudes are represented by a single on-shell diagram.
\input{figures/Onshell45}
This extends to $n$-point MHV as BCFW recursion only contains one term at each order. 
%
%
\input{figures/OnshellMHV}   
Beyond MHV, BCFW recursion decompose the amplitude into multiple on-shell diagrams. For instance, the 6-point NMHV can be expressed as
%
%
\input{figures/Onshell6}
where we recognize the 4pt MHV and 5pt MHV building blocks embedded in the 6pt NMHV diagrams, with additional BCFW bridges. We will study these BCFW terms in great details later. 

Before proceeding further, let us make two important remarks. First, if we choose adjacent BCFW shifts exclusively at each order of recursion, we only get planar on-shell diagrams; non-planar on-shell diagrams arise when non-adjacent legs are shifted. Second, the set of all on-shell diagrams relevant for the BCFW recursion relations (planar or non-planar) is a (particularly simple) subset of all leading singularities.

\subsection{From On-Shell Diagrams to Grassmannian Geometry}

On-shell diagrams also appear in combinatorics and algebraic geometry as \emph{plabic graphs} and are related to the cells of Grassmannian $G(k,n)$, where $k$ and $n$ specify the underlying helicity configuration and the number of external legs respectively. To each \emph{planar} plabic graph we can associate a cell in the \emph{positive Grassmannian} $G_{+}(k,n)$, represented by a $(k\times n)$ real matrix $C$ modulo $GL(k)$ transformations which has all main $(k\times k)$ ordered minors being positive. The plabic graph provides a particular way to parameterize the matrix via \emph{boundary measurement}. Specifically, a boundary measurement gives a set of variables $\alpha_i$'s labelling the egdes of the graph; when these variables are real with definite signs, the $C$-matrix has all main minors positive. The \emph{top cell} of $G_+(k,n)$ has maximal $k(n{-}k)$ dimensionality and is parameterized $k(n{-}k)$ of parameters; all other cells are lower-dimensional boundaries of the top cell where some parameters are sent to zero -- in the plabic graph, this corresponds to removing edges . See \cite{Arkani-Hamed:2012zlh} for more details.

As an example, we can show the same diagram from the previous section. It corresponds to a \emph{top cell} of $G_+(2,5)$, and a particular positive parametrization of the $(2\times5)$ matrix found using boundary measurements is as follows:
\input{figures/Cmatrix}
For $\alpha_k>0$ all $(2\times2)$ ordered minors are positive. 

A cell in the positive Grassmannian $G_+(k,n)$ is a space with boundaries and an example of the {\it positive geometry} \cite{Arkani-Hamed:2017tmz}. The geometry can be interpreted as a $k$-plane spanned by the rows of $C$ in $n$-dimensions subject to positivity constraints on its parameters. But it is illuminating to view the $C$-matrix instead as a collection of $n$ columns vectors $\vec{P}_i$ in the projective space $\mathbb{P}^{k{-}1}$
\begin{equation}
    C = \left(\begin{array}{cccccc} \ast & \ast & \ast & \ast & \dots & \ast \\ \ast & \ast & \ast & \ast & \dots & \ast\\ \vdots & \vdots & \vdots & \vdots & \vdots & \vdots \\ \ast & \ast & \ast & \ast & \dots & \ast
    \end{array}\right) \quad \rightarrow \quad  
    \left(\begin{array}{cccccc} \vec{P}_1 & \vec{P}_2 & \vec{P}_3 & \vec{P}_4 &\dots & \vec{P}_n
    \end{array}\right)\,,
\end{equation}
Modding out the scale from each vector %
\begin{equation}
    \vec{P}_i = \lambda_i P_i \qquad\mbox{with $\lambda_i>0$}\,.
\end{equation}
$P_i$ then describes a point in the projective space and the cell in $G_+(k,n)$ can be interpreted as as a particular configuration of $n$ points in the projective space $\mathbb{P}^{k{-}1}$. Let us consider the  $k=2$ and $k=3$  scenarios explicitly.

\paragraph{$\mathbf{k=2}$, MHV case}
The projective space is just $\mathbb{P}^1$, ie. projective line, and each $P_i$ is a point on this projective line:
\begin{equation}
    P_i = \left(\begin{array}{cc} 1 & x_i \end{array}\right)\,. \label{par}
\end{equation}
The $(2\times2)$ minor of $\vec{P}_i$ then reduces to
\begin{equation}
    (ij) = |\vec{P}_i\,\vec{P}_j| = \lambda_i\lambda_j\left|\begin{array}{cc} 1 & 1 \\ x_i & x_j\end{array}\right| = \lambda_i\lambda_j(x_j - x_i)\,.
\end{equation}
Positivity of all such ordered minors, $(ij)>0$ for $j>i$ implies an ordering of parameters $x_i$,
\begin{equation}
    x_1 < x_2 < x_3 < \dots < x_n\,.
\end{equation}
Geometrically, this corresponds to $n$ ordered points on a projective line,
%

%
\input{./figures/Line1.tex}
The boundaries of this geometry are lower dimensional cells which correspond to sending some of the consecutive minors $(i\,i{+}1)=0$. Sending one minor to zero, for example, \,$(12)=\lambda_1\lambda_2(x_2-x_1)=0 \rightarrow x_1=x_2$,   corresponds to merging two adjacent two points 
\begin{align}
    \input{./figures/Line2.tex}\label{merge12}
\end{align}
Sending another minor $(j\,j{+}1)=0$ would merge respective points and we can continue this process by probing lower and lower dimensional boundaries. We can also send the overall factor $\lambda_i\rightarrow0$ and delete the point $i$ point completely. This gives us two types of codimension-2 boundaries, 
%
%

%
\input{./figures/MHVconfig.tex}
Note that we can only remove point $i$ if $P_i$ is already identified with other $P_j$ (either $P_{i{-}1}$ or $P_{i{+}1}$), otherwise the dimensionality of the configuration would decrease by 2, rather than 1. This continues to even lower dimensional boundaries when we merge or delete more points.

\paragraph{$\mathbf{k=3}$, NMHV case} The Grassmannian geometry is  configurations of $n$ points in $\mathbb{P}^2$, the generic configuration corresponding to the $(3n-9)$-dimensional top cell are arranged as 
\input{figures/NMHVconfig}
The convexity of the configuration follows from the positivity of all  $(3\times3)$ consecutive minors. Sending consecutive minors $(i{-}1\,i\,i{+}1)=0$ places three adjacent points on the same line, for instance 
\input{figures/NMHVgeom1}
Similarly, setting another consecutive minor $(j{-}1\,j\,j{+}1)=0$, places the respective triplet of points on a line. When two overlapping minors both vanish, e.g. in the above example further setting $(234)=0$ , we get two solutions: either all points 1, 2, 3 and 4 are on the same line or points 2 and 3 coincide. 
\input{figures/NMHVgeom2}
This can be also understood algebraically. When $(123)=0$ we can express $P_3=\alpha P_1 + (1-\alpha) P_2$ and the minor $(234)$ factorizes,
\begin{equation}
    (234) = -\frac{\lambda_2}{\lambda_1}\alpha (124)\,.
\end{equation}
Now we can either set $(124)=0$ to access the first configuration or set $\alpha=0$ to access the second. Note that unlike in the MHV case where merging two points is achieved by imposing a single constraint, we need two constraints in the NMHV case. In addition we have an overlapping case where $(124)=\alpha=0$,
\begin{align}
\input{figures/NMHVconfig2}
\end{align}
which is $(3n{-}12)$-dimensional, ie. codimension-3 configuration.

\paragraph{}
We can also consider removing a point completely. If we remove a point directly by sending $\lambda_i\rightarrow0$ the dimensionality of the configuration would drop by $3$. Hence, in order to eliminate a point, we must first put this point on a line with at least two other points, then merge it with an adjacent point before finally removing it. 
\input{figures/NMHVgeom3}
Iteratively these procedures allow us to access all lower-dimensional cells of $G_{+}(3,n)$. 


%
The discussion above generalizes to higher $k$. The top cell of $G_+(k,n)$ with all consecutive minors positive corresponds to a generic convex configuration of $n$ points in $\mathbb{P}^{k{-}1}$. Any lower-dimensional cell results from moving some of these points to lower-dimensional spaces (eg. on special planes, lines , etc) or removing some of them completely, which can be achieved by iteratively sending consecutive minors to zero and keeping in mind that minors can factorize wherein we get multiple solutions. For more details see \cite{Arkani-Hamed:2009ljj,Arkani-Hamed:2012zlh}.

Going back to the plabic graphs: assessing lower-dimensional boundaries by sending one of the edge variables to zero corresponds to removing that edge from the graph. Hence, the plabic graphs provide natural \emph{stratifications} of the positive Grassmannian $G_+(k,n)$. In our $G_+(2,5)$ example, we can make minor $(12)$ vanish by setting $\alpha_2=0$ to vanish and the resulting plabic graph describes $2n{-}5$ dimensional cell in $G_+(2,5)$. 
\input{figures/EdgeRemove}
As we already know, this corresponds to merging points $1,2$ as shown in (\ref{merge12})  in the geometric picture.

\subsection{Canonical dlog forms}

For a top cell in $G_+(k,n)$ we can associate a canonical differential form
\begin{equation}
    \Omega_{n,k}^{{\rm top}} = \frac{d^{k\times n}C}{{\rm vol}[GL(k)]} \frac{1}{(12{\dots} k)(23{\dots} k{+}1)(n1{\dots}k{-}1)}\,,\label{Omegatop}
\end{equation}
where we wedge all the (differentials of) parameters of the $C$-matrix modulo $GL(k)$ redundancy. The denominator contains the product of all {\it consecutive minors} which specify the codim-1 boundaries of the top cell. The canonical form $\Omega_{n,k}$ for a lower dimensional cell is obtained by taking subsequent residues of (\ref{Omegatop}) where we set the corresponding minors to zero,
\begin{equation}
    \Omega_{n,k} = \frac{d^{k\times n}C}{{\rm vol}[GL(k)]} \frac{1}{(12{\dots} k)(23{\dots} k{+}1)(n1{\dots}k{-}1)}\Bigg|_{({\rm minors})=0}\,.\label{Omega2}
\end{equation}
Le us look at some simple examples. For $G_+(2,n)$ the canonical form for the top form is 
\begin{equation}
    \Omega_{n,k=2}^{{\rm top}} = \frac{d^{2\times n}C}{{\rm vol}[GL(2)]}\frac{1}{(12)(23)(34)\dots(n1)}.
\end{equation}
To get the canonical form associated with the cell (\ref{merge12}), we take the residue on $(12)=0$. This can be achieved by setting $x_1=x_2$ in the partial parametrization (\ref{par}):
\begin{equation}
    \Omega_{n,k=2} = \frac{d^{2\times n-1}C}{{\rm vol}[GL(2)]}\frac{1}{\lambda_1(23)(34)\dots(n2)}.
\end{equation}
Similarly, starting from the top form of $G_+(3,n)$
\begin{align}
    \Omega_{n,k=3}^{{\rm top}} &= \frac{d^{3\times n}C}{{\rm vol}[GL(3)]}\frac{1}{(123)(234)(345)\dots(n12)}\,,
\end{align}
we get for the codimension-2 cell corresponding to the second configuration in \eqref{fig:NMHVgeom2} 
\begin{align}
    \Omega_{n,k=3} &= \frac{d^{3\times n-2}C}{{\rm vol}[GL(3)]}\frac{1}{\lambda_2(134)(345)(456)\dots(n13)}\,,
\end{align}
by eliminating two degrees of freedom in vector $\vec{P}_2$. 

We can write these canonical forms explicitly using any parametrization $\{x_i\}$ of the $C$-matrix, obtaining for a $m$-dimensional cell
\begin{equation}
    \Omega_{n,k} = F(x_1,x_2,{\dots},x_m)\,dx_1{\dots}dx_m \ , \label{rational-form}
\end{equation}
where $F(x_1,x_2,{\dots},x_m)$ is a rational function. With a proper choice of parameterization, the canonical form reduces to a simple \emph{dlog form},
\begin{equation}
    \Omega_{n,k} = \frac{d\alpha_1}{\alpha_1}\frac{d\alpha_2}{\alpha_2}\dots \frac{d\alpha_m}{\alpha_m} =  d\log(\alpha_1) \wedge d\log(\alpha_2) \dots \wedge d\log(\alpha_m). \label{Omega3}
\end{equation}
making it possible to access the form for a lower-dimensional cell as a simple residue about $\alpha_i=0$. The edge variables $\{\alpha_i\}$ given by the boundary measurement of the plabic graph are one such convenient choice. Note that by explicitly parameterizing the $C$-matrix, we have assumed a particular gauge-fixing of $GL(k)$ in (\ref{Omega2}). We have traded the manifest $GL(k)$ gauge invariance of (\ref{Omega2}) for exposing that all singularities of the form $\Omega_{n}^{k}$ are \emph{logarithmic} in (\ref{Omega3}).

\subsection{Dual formulation}

There is a direct connection between on-shell diagrams we defined in (\ref{onshelldiag}) and the canonical forms for the cells in the positive Grassmannian. For any on-shell diagram in a given theory, the superfunction as computed by gluing 3-pt amplitudes has the following Grassmannian integral representation\footnote{All integrals are to be evaluated as contour integrals, i.e. 
\begin{equation}
    \int \frac{dx}{x} \delta(x-x_0) = \frac{1}{x_0}.
\end{equation}},
\begin{equation}
    {\cal F}_{n,k}(\lambda,\tilde\lambda,\tilde\eta) = \int \Omega_{n,k}\,\,\delta^{k\times 2}(C{\cdot} \widetilde{\lambda})\,\delta^{2\times(n-k)}(C^\perp{\cdot} \lambda)\,\delta^{k\times \mathcal N}(C{\cdot} \widetilde{\eta})\,. \label{dual}
\end{equation}
The specific choice of the differential form $\Omega$ is theory-dependent. For the ${\cal N}=4$ SYM theory, this is a canonical dlog form (\ref{Omega3}). The differential form $\Omega$ is also known for ${\cal N}<4$ SYM \cite{Arkani-Hamed:2012zlh} and supergravity \cite{Herrmann:2016qea,Heslop:2016plj}. 
The $\delta$-functions in (\ref{dual}) play an important role as they relate the edge variables $\alpha_i$ and the kinematics $\lambda_i$, $\widetilde{\lambda}_i$, $\widetilde{\eta}_i$. The bosonic $\delta$-functions have a clear geometric interpretation. Thinking of both $\lambda$, $\widetilde{\lambda}$ as 2-planes in $n$-dimensions:
\begin{equation}
    \lambda = \left(\begin{array}{cccc}\lambda_1^{\alpha=1} & \lambda_2^{\alpha=1} & \dots & \lambda_n^{\alpha=1}\\
    \lambda_1^{\alpha=2} & \lambda_2^{\alpha=2} & \dots & \lambda_n^{\alpha=2}
    \end{array}\right), \qquad 
     \widetilde{\lambda} = \left(\begin{array}{cccc}\widetilde{\lambda}_1^{\dot{\alpha}=1} & \widetilde{\lambda}_2^{\dot{\alpha}=1} & \dots & \widetilde{\lambda}_n^{\dot{\alpha}=1}\\
    \widetilde{\lambda}_1^{\dot{\alpha}=2} & \widetilde{\lambda}_2^{\dot{\alpha}=2} & \dots & \widetilde{\lambda}_n^{\dot{\alpha}=2}
    \end{array}\right)\,.
\end{equation}
The first set of $\delta$-functions constrain the $k$-plane $C$ to be orthogonal to the 2-plane $\widetilde{\lambda}$, in other words, the $(n-k)$-plane $C^{\perp}$ contains $\widetilde{\lambda}$; the second set of $\delta$-functions says $C$ contains the 2-plane $\lambda$. Together these $\delta$-functions imply the orthogonality of 2-planes $\lambda$ and $\widetilde{\lambda}$, 
\input{figures/Perp2Planes}
which is precisely the statement of momentum conservation in the $n$-dimensional (particle) space. 
\begin{equation}
    \delta^4(P) \equiv \delta^{2\times2}(\lambda\cdot\widetilde{\lambda}). \label{mom}
\end{equation}
%
Furthermore, with $\lambda \subset C$, the Grassmann $\delta$-functions in (\ref{dual}) impose the super-momentum conservation (along with additional constraints on $\widetilde{\eta}$ for $k>2$)
\begin{equation}
    \delta^{2\mathcal N}(Q)\equiv \delta^{2\times \mathcal N}(\lambda\cdot\widetilde{\eta})\,.
\end{equation}
Out of the $2n$ bosonic $\delta$-functions, $(2n-4)$ constraints are available for localizing the integral. We can always use part of the $\delta(C^\perp{\cdot}\lambda)$ constraints to fix the first two rows of the $C$-matrix to $\lambda$, 
%
\begin{equation}
    C = \left(\begin{array}{ccccc} \lambda_1^{(1)}& \lambda_2^{(1)} & \lambda_3^{(1)}& \dots & \lambda_n^{(1)}\\
    \lambda_1^{(2)}& \lambda_2^{(2)} & \lambda_3^{(2)}& \dots & \lambda_n^{(2)}\\ \ast & \ast & \ast & \dots & \ast \end{array}\right)  = \left(\begin{array}{c} \lambda \\ C^\ast \end{array}\right)\,.\label{Csol}
\end{equation}
then factor out the momentum and super-momentum conserving explicitly from the remaining $\delta$-functions,
\begin{equation}
    \delta^{k\times2}(C\cdot\widetilde{\lambda}) \delta^{k\times \mathcal N}(C\cdot\widetilde{\eta}) = \delta^4(P)\delta^{2\mathcal N}(Q) \times \delta^{(k-2)
    \times 2}(C^\ast\cdot\widetilde{\lambda}) \delta^{ (k-2)\times\mathcal N}(C^\ast\cdot\widetilde{\eta}).
\end{equation}
The residual constraints only involving $(k{-}2)\times n$ matrix $C^\ast$. Depending on the dimensionality of the Grassmannian cell, the full integral will evaluate to one the following 
\begin{itemize}
    \item if $m=2n-4$: ordinary superfunctions. All degrees of freedom of the $C$-matrix are precisely fixed by the $\delta$-function constraints. This is exactly the case of on-shell diagrams that appear in the BCFW recursion relations, and in general all leading singularities fall into this category. 
    \item if $m>2n{-}4$: differential forms with unfixed parameters. The remaining degrees of freedom can be interpreted as components of loop momenta $\ell_i$ not determined by cuts.
    \item if $m<2n{-}4$: singular limits of superfunctions, where the external kinematics are subject to additional constraints 
\end{itemize}

Let us go back to our example of on-shell diagram (\ref{onshelldiag}). In ${\cal N}=4$ SYM this diagram evaluates to
\begin{equation}
    {\cal F}_{5,2} = \frac{1}{J}\,\prod_j A_{3pt} = \int \Omega_{5,2}\,\delta^{2\times 2}(C\cdot \widetilde{\lambda})\,\delta^{3\times 2}(C^\perp\cdot \lambda)\,\delta^{2\times 4}(C\cdot \widetilde{\eta})\,.
    \label{example1}
\end{equation}
The $C$-matrix is a 2-plane in $5$-dimensions which must contain another 2-plane $\lambda$. We can set  $C=\lambda$ trivially. The other two $\delta$-functions reproduce momentum and supermomentum conservation and we get as expected
\begin{equation}
    {\cal F}_{5,2} = \frac{\delta^{2\times 2}(\lambda\cdot\widetilde{\lambda})\delta^{2\times 4}(\lambda\cdot\widetilde{\eta})}{\la12\ra\la23\ra\la34\ra\la45\ra\la51\ra}\,.
   \label{example2}
\end{equation}
Note that if we were to use the edge variables provided by the plabic graph, we would not get the solution in the form of $C=\lambda$. Instead we would find, for instance,
\begin{equation}
    C = \left(\begin{array}{cccccc} 1 & 0 & \frac{\la14\ra}{\la12\ra}& \frac{\la 15\ra}{\la 12\ra}& \dots & \frac{\la 1n\ra}{\la 12\ra} \\
    0 & 1 & \frac{\la 24\ra}{\la 21\ra} & \frac{\la 25\ra}{\la 21\ra} & \dots & \frac{\la 2n\ra}{\la 21\ra}\end{array}\right)\,.
\end{equation}
this is related to $C=\lambda$ by a $GL(2)$ transformation. The two matrices represents the same point in the Grassmannian, only parameterized with different gauge-fixing schemes. The expression (\ref{example1}) obviously does not depend on the choice of the $GL(2)$ fixing and we always get the same result.

\subsection{Cells for NMHV Tree-level Amplitudes}
Now we turn our attention to NMHV tree-level amplitudes, which will be our main focus throughout this paper. The Grassmannian cells appearing in the BCFW construction of tree amplitudes are always $(2n-4)$-dimensional. Since in the NMHV case, the top cell of $G_+(3,n)$ has the dimensionality $3n-9$, the relevant cells for the amplitude must be of co-dimension $n-5$. In other words, they are obtained from the top cell by imposing $n-5$ constraints (on consecutive minors).

The 5-point case is trivial. The BCFW representation contains only a single term, which is a parity conjugate of the 5-pt MHV amplitude. There is no constraint to be imposed -- the single BCFW term directly corresponds to the top cell of $G(3,5)$ and the generic convex configuration of five points in $\mathbb{P}^2$. The first interesting case appear at 6-point where the BCFW representation of the amplitude, using the $(61)$ shift for instance, consists of three on-shell diagrams (\ref{6ptNMHV}). Each diagram represents an 8-dimensional configuration in $G_{+}(3,6)$ with one constraint impose on a configuration of six points in $\mathbb{P}^2$,
\input{figures/NMHVgeom5}
The first term has minor $(123)=0$ as reflected by points 1, 2 and 3 lying on the same line in the picture. Solving the $\delta$-function constraints we get the form of the $C$-matrix,
\begin{equation}
    C = \left(\begin{array}{cccccc} \lambda_1^{(1)} & \lambda_2^{(1)} & \lambda_3^{(1)} & \lambda_4^{(1)} & \lambda_5^{(1)} & \lambda_6^{(1)} \\
    \lambda_1^{(2)} & \lambda_2^{(2)} & \lambda_3^{(2)} & \lambda_4^{(2)} & \lambda_5^{(2)} & \lambda_6^{(2)} \\ 
    0 & 0 & 0 & [56] & [64] & [45] \end{array}\right)\,. \label{NMHV6pt_example}
\end{equation}
The superfunction (\ref{dual}) then takes a particularly simple form, 
\begin{equation}
{\cal R}_1 = \frac{\delta^4(P)\delta^8(Q)\delta^4([56]\widetilde{\eta}_4 + [64]\widetilde{\eta}_5 + [45]\widetilde{\eta}_6)}{s_{123} \la12\ra\la23\ra[45][56]\la 1|23|4]\la 3|45|6]}\,,
\end{equation}
where we denoted $\la1|23|4] \equiv \la12\ra[24] + \la13\ra[34]$. All other expressions are related by cyclic shifts ${\cal R}_1\rightarrow {\cal R}_{i{+}1}$ where we just relabel $k\rightarrow k+i$. The 6-point tree-level amplitude is then given by
\begin{equation}
    {\cal A}_6^{(3)} = {\cal R}_1 + {\cal R}_3 + {\cal R}_5\,. \label{SixPoint1}
\end{equation}
If we perform the BCFW recursion relation with the shift $(12)$, we get a different set of on-shell diagrams. They are are constructed in the same way, but the BCFW bridge attached is now $(12)$, rather than $(n1)$. The set of three Grassmannian configurations is following,
\input{figures/NMHVgeom6}
The superfunction for each configuration can be again obtained from ${\cal R}_1$ by a cyclic shift,
\begin{equation}
     {\cal A}_6^{(3)} = {\cal R}_2 + {\cal R}_4 + {\cal R}_6\,. \label{SixPoint2}
\end{equation}
The equality of (\ref{SixPoint1}) and (\ref{SixPoint2}) is guaranteed by the Global Residue Theorem (GRT) for the superfunction associated with the top cell of $G_+(3,6)$ (which has an extra parameter $z$ and vanishes for $z\rightarrow\infty$). 

Going to higher points  BCFW recursion expands the amplitude as a sum of two types of on-shell diagrams:
\input{./figures/BCFWNMHV1.tex}

The first type of on-shell diagrams corresponds to the factorizations of the NMHV amplitude into two MHV amplitudes; each MHV blob corresponds to a single on-shell diagram; gluing them with the BCFW bridge makes an NMHV on-shell diagram. It is not hard to see every diagram thus constructed corresponds to a Grassmannian configuration where all $n$ points are localizes on three lines:
\input{./figures/BCFWNMHV3.tex}
This is indeed $2n-4$ dimensional configuration whose $C$-matrix has vanishing minors:
\begin{equation}
   (123)=(234)=\dots=(i{-}2\,i{-}1\,i)=(i{+}1\,i{+}2\,i{+}3)=\dots=(n{-}3\,n{-}2\,n{-}1)=0 
\end{equation}
while all other minors $(ijk)>0$. Note that none of the points are merged with another or removed with these constraints.

The second term in (\ref{NMHVconf}) corresponds to a \emph{sum} of on-shell diagrams which arise from the factorization of the amplitude into a $(n{-}1)$-pt NMHV and a 3-point $\overline{\rm MHV}$ amplitude. It can also be interpreted as adding particle $n$ to the $(n{-}1)$-pt NMHV amplitude via an $k$-preserving inverse-soft factor. Suppose we have found the diagrammatic expansion of the $(n{-}1)$-point NMHV amplitude via recursion (\ref{NMHVconf}), then adding the inverse-soft factor corresponds to adding point $n$ between $n{-}1$ and $1$ for every Grassmannian configuration underlying the $(n{-}1)$-point NMHV amplitude, which involve $n{-}1$ points distributed on three lines with special point $1$ on the intersection of two lines
\input{./figures/BCFWNMHV4.tex}
The resulting configurations have points $n$ and $n{-}1$ on the same line. They complement the first set of on-shell diagrams (\ref{BCFWNMHV3}) where point $n{-}1$ is on a different line than point $n$. 

Indeed if we consistently use the same shift on the lower-point NMHV amplitudes (e.g. for the subamplitude $A(I,\hat{1},2,3,{\dots},i)$ on the factorization channel we use $(I1)$ shift in the recursion), we obtain a closed-form solution for general $n$-point NMHV amplitudes as a sum over a special set of configurations
\begin{align}
    \mathcal{A}^\text{NMHV}_n=\sum_{i,j}\hspace{0.5cm}
\input{./figures/BCFWNMHV2.tex} 
    \label{BCFWNMHV2}
\end{align}
where all $n$ points are localized on three lines and each line has at least two points on it. Each configuration is characterized by 3 labels: the first label $1$ (as $1$ in the $(n1)$ shift used above) specifying the ``center'' of the configuration,  the other two labels $i$ and $j$ specifying the ``boundary'', where $i\geq 2$, $i{+}2\leq j\leq n{-}1$ ensures that at least two points are on each line. The Grassmannian cell (and also the permutation) can be easily read off from the configuration:
\begin{align}
    &(123)=(234)={\dots}=(i{-}2\,i{-1}\,i)=(i{+}1\,i{+}2\,i{+}3)={\dots}(j{-}2\,j{-}1\,j)=\nonumber\\\hspace{1cm}&=(j{+}1\,j{+}2\,j{+}3)={\dots}(n{-}1\,n\,1)=0\qquad
    \mbox{while all others $(ijk)>0$}.
\end{align}

The corresponding superfunction ${\cal F}$ is
\begin{equation}
    {\cal R}_{1,i{+}1,j{+}1} = \input{./figures/BCFWNMHV2.tex}  =  \frac{\delta^4(P)\delta^8(Q)}{\la12\ra\la 23\ra\dots \la n1\ra}\times R[1,i,i{+}1,j,j{+}1]\,,\label{Rinv}
\end{equation}
where we have used the first points on each line in the general configuration (\ref{BCFWNMHV2}) to label the expression. We also introduce the R-invariant in the momentum twistor space
\begin{equation}
    R[a,b,c,d,e] = \frac{(\eta_a\la bcde\ra + \eta_b \la cdea\ra + \eta_c\la deab\ra + \eta_d\la eabc\ra +\eta_e\la abcd\ra)^4}{\la abcd\ra\la bcde\ra\la cdea\ra\la deab\ra\la eabc\ra}\,,
\end{equation}
where $\eta_k$ are the momentum twistor Grassmann variables \cite{Arkani-Hamed:2009nll,Mason:2009qx}. This representation makes both the superconformal and the dual superconformal symmetries (and hence the Yangian symmetry) manifest. For our purpose, we rewrite the $R$-invariant in the momentum space,
\begin{equation}
\label{eq:Rinv}
    R[1,i,i{+}1,j,j{+}1] = \frac{\la i\,i{+}1\ra\la j\,j{+}1\ra \cdot  \delta^4(\Xi_{1,i{+}1,j{+}1})}{P_2^2\la 1|P_1P_2|j\ra\la 1|P_1P_2|j{+}1\ra\la 1|P_3P_2|i\ra\la 1|P_3P_2|i{+}1\ra}\,,
\end{equation}
where we denoted
\begin{equation}
    P_1 = p_2+\dots+p_i,\qquad P_2 = p_{i{+}1}+\dots+p_j,\qquad P_3 = p_{j{+}1}+\dots+p_n\,,
\end{equation}
such that $P_1+P_2+P_3+p_1 = 0$. Pictorially $P_1$, $P_2$, $P_3$ are the collections of points on each of the three lines:
%
%
\input{./figures/ThreeLines.tex}
The argument of the super delta function $\delta^4(\Xi_{1,i{+}1,j{+}1})$ is given by
\begin{equation}
    \Xi_{1,i{+}1,j{+}1} \equiv \sum_{k\in P_2} \la k|P_2P_3|1\ra \widetilde{\eta}_k + \sum_{j\in P_3} P_2^2 \la 1j\ra \widetilde{\eta}_j\,. \label{Xi}
\end{equation}
Note that this formula does not depend on the ordering in sets $P_1$, $P_2$ and $P_3$ on the respective lines. This will be very important in the next section when we discuss non-adjacent BCFW recursion. As might be evident, the expression (\ref{Xi}) only depends on Grassmann variables $\widetilde{\eta}$s from lines $P_2$ and $P_3$. Thus we can use super-momentum conservation to rewrite it in an equivalent form when $P_1\leftrightarrow P_3$ (reflection of the picture),
\begin{equation}
    \Xi_{1,i{+}1,j{+}1} \equiv \sum_{k\in P_2} \la k|P_2P_1|1\ra \widetilde{\eta}_k + \sum_{j\in P_1} P_2^2 \la 1j\ra \widetilde{\eta}_j\,. \label{Xi2}
\end{equation}

The set of all BCFW terms generated by adjacent shifts is a subset of all \emph{planar} on-shell diagrams of dimensionality $2n-4$ (which are leading singularities). For a general leading singularity, the $(2n-4)$-dimensional cell in $G_+(3,n)$ corresponds to a configuration of $n$ points in $\mathbb{P}^2$ located on five lines, 
\input{figures/GeneralNMHV1}
Any term which appears in the context of adjacent BCFW recursion is a special case, where two of the lines have exactly two points and we get the configuration discussed earlier,
\input{figures/GeneralNMHV2}
%
%

\paragraph{Recap:}
Let us close the section with a summary of a few important points in the discussion
\begin{itemize}
    \item Each BCFW term is a sum of on-shell diagrams. 
    \item Each on-shell diagram is corresponds to a $2n{-}4$ dimensional cell in $G_+(k,n)$ with a super-function that can be calculated using (\ref{dual}).
    \item Each cell in $G_+(k,n)$ is characterized by a configuration of $n$ points in $\mathbb{P}^{k{-}1}$ with only consecutive constraints.
    \item For MHV, $k=2$, the amplitude is given by the top cell of $G_+(2,n)$ which describes $n$ ordered points on a projective line $\mathbb{P}^1$, and the superfunction  reproduces a famous Parke-Taylor formula.  
    \item For NMHV, $k=3$, the amplitude is given by a collection of co-dimension $(n{-}5)$ cells in $G_+(3,n)$. These cells describe configurations of $n$ points in $\mathbb{P}^2$ localized on three lines and the superfunctions are the $R$-invariants with special labels.
\end{itemize}

\noindent It is worth repeating that having the Grassmannian configurations, we can easily reconstruct the $C$-matrix and calculate the associated superfunction (\ref{dual}). Therefore, we focus mainly on these geometries here.

\section{Holomorphic expressions for BCFW terms}
\label{sec:holoform}

The $\delta$-functions constraints in (\ref{dual}) provide a map between the Grassmannian space and the kinematic space through $C\cdot \widetilde{\lambda}=C^\perp\cdot \lambda=0$. In this section we will recast the Grassmannian geometries of $C$-matrices in the space of $\lambda$ (we refer to it as holomorphic space). We will see that the interpretation of the geometry in the kinematic space lead to new, intuitive representations of BCFW terms.   

As a trivial example of the $\lambda$-geometry, consider the case of $k=2$ MHV geometry. In this case $C=\lambda$ we identify the $C$-matrix with the $2$-plane spanned by $\lambda$, and the positive Grassmannian geometry can be directly interpreted as the $\lambda$-geometry, that is, $\vec{P}_k=\lambda_k$ -- the column vector $\vec{P}_k$ for each point on the projective line $\mathbb{P}^1$ is equal to the kinematic variable $\lambda$. 
\input{figures/LineSing}
Merging points 1 and 2 sets the minor $(12)=0$ and this is the same as $\la12\ra\rightarrow0$. Removing the point 2 from the picture corresponds to sending $\lambda_2\rightarrow0$ and so on.

For $k>2$ the connection between the $C$-matrix and kinematical constraints on $\lambda$, $\widetilde{\lambda}$ is not straightforward. We need to solve the conditions $C\cdot \widetilde{\lambda}= C^\perp\cdot \lambda=0$, express the $C$ matrix using $\lambda$, $\widetilde{\lambda}$ and evaluate minors of the $C$ matrix. Let us look at a 6-pt NMHV configuration associated with discussed earlier:
\input{figures/NMHVgeom7}
This is a 8-dimensional cell with $(123)=0$. Solving for $C$ from the delta functions we get explicit form (\ref{NMHV6pt_example}). We can evaluate remaining minors:
\begin{equation*}
    (234) = \la 23\ra [56],\  (345) = \la3|45|6], \  (456) = s_{123},\  (561) = \la 1|23|4],\  (612) = \la 12\ra [45]\,.
\end{equation*}
Let us look at $(234)=0$, which has two solutions (\ref{fig:NMHVgeom2}): $1,2,3,4$ on the same line or merging points $2,3$. While the first one sends $[56]=0$, the latter corresponds to $\la23\ra=0$. Similarly, $(612)=0$ puts points $6,1,2,3$ on a line ($[45]=0$) or merges $1,2$ ($\la 12\ra=0$). We can see that merging points on the $1,2,3$ line gives $\la12\ra=0$ or $\la23\ra=0$ which resembles the $C=\lambda$ correspondence for the $k=2$ geometry.
The other boundaries are more complicated and there is no obvious pattern in these singularities. For instance, putting $3,4,5$ on a line by $(345)=0$ sends $\la3|45|6]=0$ which is hard to interpret. Furthermore the poles are no longer holomorphic $\la ab\ra=0$ like in the MHV case, ie. they generally depend on both $\lambda$ and $\widetilde{\lambda}$. 

Nevertheless, our intention is to make a direct connection between the Grassmannian geometry and the kinematical space. The motivation comes from the analysis of the BCFW term (\ref{BCFWNMHV3}), which originates as the product of two MHV amplitudes, 
\begin{align}
   &\begin{tikzpicture}[scale=1, baseline={(0,0cm)}]
    \draw(1,0)--(0,0) node [at end, left] {$n-1$};
    \draw(1,0)--(1,-1) node [at end, below] {$i+1$};
    \draw(1,0)--(0.25,0.75) node [at end, above left] {$\hat{n}$};
    \draw(1,0)--(2.25,0)
    node[pos=0.5,above]{$I$};
    \draw(2.25,0)--(3.25,0) node [at end, right] {$2$};
    \draw(2.25,0)--(3,0.75) node [at end, above right] {$\hat{1}$};
    \draw(2.25,0)--(2.25,-1) node [at end, below] {$i$};
    \node[rotate=45] at (3,-0.75) {\large$\cdots$};
    \node[rotate=-45] at (0.25,-0.75) {\large$\cdots$};
    \draw[black,fill=white] (1,0) circle (1.25ex);
    \node at (1,0) {2};
    \draw[black,fill=white] (2.25,0) circle (1.25ex);
    \node at (2.25,0) {2};
    \end{tikzpicture} = A_{k{=}2}^{\rm tree}(i{+}1,i{+}2,{\dots},n{-}1,\hat{n},I)\times \frac{1}{(P_2+p_n)^2}\nonumber\\[-22pt]
    &\hspace{9.5cm} \times A_{k{=}2}^{\rm tree}(I,\hat{1},2,{\dots},i{-}1,i)\,. \label{term1}
\end{align}
Here we denote $P_1=p_2+\dots+p_i$ and we also define $P_2=p_{i{+}1}+\dots+p_{n{-}1}$. Momentum conservation implies $P_1+P_2+p_1+p_n=0$. The BCFW shift parameter is fixed to
\begin{equation}
    z = \frac{(P_1+p_1)^2}{\la 1|P_1|n]}\,,
\end{equation}
and the shifted spinors $\lambda_{\hat{n}}$, $\widetilde{\lambda}_{\hat{1}}$ and internal momentum $p_I$ are given by
\begin{equation}
    \lambda_{\hat{n}} = \frac{\la 1|P_1P_2}{\la 1|P_1|n]},\qquad \widetilde{\lambda}_{\hat{1}} = \frac{P_1P_2|n]}{\la 1|P_1|n]},\qquad
    p_I = \frac{1}{\la 1|P_1|n]}\left(P_2|n]\right)\left(\la 1|P_1\right).
\end{equation}
In this notation the spinor $\la 1|P_1P_2$ is equal to
\begin{equation}
\la 1|P_1P_2 = \sum_{i\in P_1}\sum_{j\in P_2} \la 1i\ra[ij] \lambda_j\,.
\end{equation}
Note that the division of $p_I$ into $\lambda_I$ and $\widetilde{\lambda}_I$ is ambiguous. We see that the BCFW term (\ref{term1}) is the product of two MHV amplitudes with an additional pole $\frac{1}{(P_2+p_n)^2}$. Therefore, we can modify the Grassmannian configuration for (\ref{BCFWNMHV3})
as
\input{figures/ThreeLines1}
where we added two extra points $\hat{n}$ and $I$ in the picture. Note these points originate as certain special momenta $p_{\hat{n}}$ and $p_I$ but now we associate them with points in the Grassmannian geometry. 

Let us take a closer look at the geometry: we have now points $1,2,\dots,i,I$ on the first line, points $I,i{+}1,\dots,n{-}1,\hat{n}$ on the second line and points $\hat{n},n,1$ on the third line. It is obvious that the two MHV amplitudes in (\ref{term1}) correspond to configurations of points on the first two lines in (\ref{graph1}), but it is not that clear what to do with the third line and $1/(P_2+p_n)^2$ pole. Let us make the connection precise now.
 
We start with the expression for the ${\cal R}_{1,i{+}1,j{+}1}$ and rewrite it as  
\begin{align}
    &{\cal R}_{1,i{+}1,n}\nonumber\\ &\hspace{1cm}=\frac{\widetilde{\Delta}_{1,i{+}1,n}}
    {\la 12\ra ... \la i{-}1\,i\ra\la i{+}1\,i{+}2\ra ... \la n{-}2\,n{-}1\ra {\la i|P_2|n]}\la i{+}1|P_2|n] { \la n{-}1|P_2P_1|1\ra} {(P_2{+}p_n)^2}{P_2^2}}\,,
\end{align}
where
\begin{align}
    \widetilde{\Delta}_{1,i{+}1,n}=\delta^4(P)\,\delta^8(Q)\,\delta^4(\widetilde{\Xi}_{1,i{+}1,n})\,,
\end{align}
and the reduced argument of the delta function is
\begin{equation}
\widetilde{\Xi}_{1,i{+}1,n} = \sum_{k\in P_2}\la k|P_2|n]\widetilde{\eta}_k + P_2^2\widetilde{\eta}_n = \frac{1}{\ang{1n}}\, \Xi_{1,i{+}1,n}\,.
\end{equation}
Looking at the denominator we can now clearly identify Parke-Taylor factors for MHV amplitudes on two of the lines which now also involve points $\hat{n}$ and $I$,
\begin{equation}
   PT(1,2,\dots,i{-}1,i,I)\times PT(i{+}1,i{+}2,\dots,\hat{n},I)\,,
\end{equation}
where the Parke-Taylor factors are defined as usual,
\begin{equation}
    PT(a_1,a_2,\dots a_m) = \frac{1}{\la a_1a_2\ra\la a_2a_3\ra\dots \la a_m a_1\ra}\,.
\end{equation}
Note that all the mixed poles come from \emph{holomorphic} poles involving $\lambda_I$ and $\lambda_{\hat{n}}$,
%
\begin{equation}
    {\la i\,I\ra \doteq \la i |P_2|n]},\quad {\la i{+}1\,I\ra \doteq \la i{+}1 |P_2|n]},\quad {\la n{-}1\,\hat{n}\ra \doteq \la n-1|P_2P_1|1\ra},\quad {\la \hat{n}\,I\ra \doteq P_2^2 }\,.
\end{equation}
We have ignored any factor of $\la 1|P_1|n] = - \la 1|P_2|n]$ to some power in the above equation; as noted before, the splitting $p_I$ into $\lambda_I$ and $\widetilde{\lambda}_I$ is ambiguous which produces some particular power of $\la 1|P_1|n] = - \la 1|P_2|n]$ from the Jacobian. This will be accounted for in the end. Furthermore, there is a factor $1/(P_2+p_n)^2$ which does not originate from either of the Parke-Taylor factors, but can be interpreted as 
\begin{equation}
    {\la n\,\hat{n} \ra = \la 1|P_1P_2|n\ra = (P_2+p_n)^2 \la 1n\ra\,.}
\end{equation}
In fact, noticing that $\la 1\,\hat{n}\ra = \la 1n\ra \la 1|P_1|n]$ , we can easily check all the remaining factors conspire to make another Parke-Taylor factor for the MHV amplitude living on the third line with points $1,n,\hat{n}$:
\begin{equation}
    \frac{1}{\la 1n\ra\la n\,\hat{n}\ra\la \hat{n}1\ra}\,.
\end{equation}

Taking everything together we conclude that the ${\cal R}_{1,i{+}1,n}$ can be rewritten as the product of three Parke-Taylor factors living on the three lines that form the Grassmannian geometry
\begin{equation}
   {\cal R}_{1,i{+}1,n} = PT(1,\dots,i,I_1)PT(I_1,i{+}1,\dots,n{-}1,I_2)PT(I_2,n,1)\times\la1|P_1p_n|1\ra^3\widetilde{\Delta}_{1,i+1,n}\,, \label{R1}
\end{equation}
where we have introduced two points $I_1$, $I_2$, instead of $I$, $\hat{n}$, with fixed form for their $\lambda$-spinors,
\begin{equation}
    \lambda_{I_1} = P_2|n],\qquad \lambda_{I_2} = \la1|P_1P_2\,.
\end{equation}
Note that in the formula (\ref{R1}) we have a product of three Parke-Taylor factors, rather than the full MHV amplitudes. So far we have merely rewritten ${\cal R}_{1,i{+}1,n}$ in a particular way. But, importantly, there is a clear pattern in the formula (\ref{R1}) that immediately generalize to the more general R-invariants ${\cal R}_{1,i{+}1,j{+}1}$, for which the interpretation using BCFW shifts is absent. 

In the general configuration we have $n$ points living on all three lines,
\begin{equation}
\label{ThreeLines2}
    \input{figures/ThreeLines2}
\end{equation}
and ${\cal R}_{1,i{+}1,j{+}1}$ is given by the product of Parke-Taylor factor on three lines, with labels indicated in the picture,
\begin{align}
{\cal R}_{1,i{+}1,j{+}1} = &PT(1,{\dots},i,I_1)PT(I_1,i{+}1,{\dots},j,I_2)PT(I_2,j{+}1,{\dots},n,1)\nonumber\\
&\hspace{7cm}\times\la1|P_1P_3|1\ra^3 \Delta_{1,i{+}1,j{+}1}\,, \label{R2}
\end{align}
where $P_1=p_2+{\dots}+p_i$, $P_2=p_{i{+}1}+{\dots}+p_j$ and $P_3=p_{j{+}1}+{\dots}+p_n$. The auxiliary spinors $\lambda_{I_1}$, $\lambda_{I_2}$ are given by 
\begin{equation}
    \lambda_{I_1} = \la 1|P_3P_2, \qquad\lambda_{I_2} = \la 1|P_1P_2\,, \label{spinors}
\end{equation}
and $\Delta_{1,i{+}1,j{+}1}$ is just given by the delta functions in (\ref{Rinv}),
\begin{equation}
 \Delta_{1,i{+}1,j{+}1} = \delta^4(P)\,\delta^8(Q)\,\delta^4(\Xi_{1,i{+}1,j{+}1})\,.
\end{equation}

Our expression (\ref{R2}) reproduces (\ref{Rinv}). The salient feature of the new formula is that all mixed poles in ${\cal R}_{1,i{+}1,j{+}1}$ appear as simple holomorphic poles involving $\lambda_{I_1}$ and $\lambda_{I_2}$. The introduction of points $I_1$, $I_2$ prime the original configuration of $n$ points in Grassmannian cell $G_+(3,n)$ for a kinematical interpretation. Merging two points, which corresponds to sending certain consecutive minor of the $C$-matrix to zero, turns into statement about $\la ab\ra=0$, ie. $\lambda_a\sim\lambda_b$ for two points $a,b$ on one of the three lines with points $1,2,\dots,i,I_1$ and $I_1,i{+}1,\dots,j,I_2$ and $I_2,j{+}1,\dots,n,1$. This is trivially true for the points $1,\dots,n$ but quite remarkably this also extends to all merging involving points $I_1$ and $I_2$ which lead to more complicated non-holomorphic poles, 

\input{./figures/ThreeLinesSing.tex}
As an example, let us look at the 8-point configuration ${\cal R}_{1,4,7}$.
\input{./figures/Example_R147.tex}
Following the above procedure we can build the superfunction from three Parke-Taylor factors
\begin{align}
   {\cal R}_{1,4,7} &= PT(1,2,3,I_1)PT(I_1,4,5,6,I_2)PT(I_2,7,8,1)\cdot \la1|(23)(78)|1\ra^3\Delta_{1,4,7}\\
   &\hspace{-1cm}= \frac{\Delta_{1,4,7}}{\la 12\ra\la 23\ra\la 3|(456)(78)|1\ra\la 1|(78)(56)|4\ra\la 45\ra\la 56\ra\la 1|(23)(45)|6\ra s_{456} \la 1|(23)(456)|7\ra\la 78\ra\la 81\ra}\nonumber
\end{align}
with 
\begin{equation}
   \lambda_{I_1} = \la 1|(78)(456)\, ,\qquad \lambda_{I_2} = \la 1|(23)(456)\,. \label{eq:extra-spinors}
\end{equation}

As a side note, while all we need is the $\lambda$-part of $I_1$, $I_2$ in the context of Yang-Mills theory, for the purpose of  extending the same positive geometry picture to other theories
we may try to upgrade them to full momenta. To do this, we have to ensure that momentum conservation is respected on each of the three lines. First, we can determine the $\widetilde{\lambda}$s for $I_1$ and $I_2$ uniquely from the momentum conservation, 
\begin{equation}
    \widetilde{\lambda}_{I_1} = 
    \frac{\la 1|P_1}{\la 1|P_1P_3|1\ra}, \qquad \widetilde{\lambda}_{I_2} = \frac{\la 1|P_3}{\la 1|P_3P_1|1\ra}\ ,
\end{equation}
where we can again see the normalization factor $\la 1|P_1P_3|1\ra$. In this case we get a momentum conservation $p_{I_1}+p_{I_2} + p_{i{+}1}+\dots+p_j=0$. On the other two lines, momentum $p_1$ is split into two parts $p_1=p_1^{a}+p_1^b$, where both momenta share the same $\lambda_1$ but differ in $\widetilde{\lambda}$ parts such that momentum conservation $-p_{I_2}+P_3+p_1^a=0$ and $-p_{I_1}+P_1+p_1^b=0$ is respected. 

\section{Non-adjacent BCFW recursion and non-planar positive geometry}
\label{sec:nonadjBCFW}

The main goal of this paper is to explore the space of non-planar on-shell diagrams that arise in BCFW recursion relations with non-adjacent shifts in the context of $\mathcal N=4$ SYM. We consider a non-adjacent BCFW shift $(k1)$,
\begin{equation}
    {\lambda}_{\hat{k}} = \lambda_k + z\lambda_1,\qquad 
   \tilde\lambda_{\hat{1}} = \tilde {\lambda}_{1} - z \tilde \lambda_k, \qquad \widetilde{\eta}_{\hat{1}} =  \widetilde{\eta}_1 + z \eta_k\,,
\end{equation}
which gives rise to an expansion of tree-level amplitudes in terms of building blocks that are superconformal invariant.
%
\input{figures/NonBCFW2.tex}
%
However, for general $k$, the dual super conformal invariance of the $\mathcal N=4$ SYM amplitude is broken in the individual terms as the cyclic ordering of points is spoiled. Nonetheless, we can still represent these terms by \emph{non-planar on-shell diagrams} %
\input{figures/NonBCFW1.tex}
where each blob can recursively expressed in terms trivalent plabic graphs. For example, one of the non-planar on-shell diagrams in the BCFW expansion of 6pt NMHV amplitude using $(51)$ shift is 
\input{figures/NonPlanarOnShell2}
%

While on-shell diagrams that are planar correspond to cells in the positive Grassmannian $G_+(k,n)$,
an arbitrary non-planar diagram corresponds to a certain cell in (the non-positive part of) $G(k,n)$. Specifically, 
For each non-planar on-shell diagram we can construct the $C$-matrix via the boundary measurement. The rules are the same as in the planar case: label the edge variables (here face variables can not be used), choose a perfect orientation and calculate the entries of the $C$-matrix. For the example above we get 
\begin{equation}
C =   \left(
\begin{array}{cccccc}
 1 & 0 & 0 & -\alpha_2\alpha_5\alpha_6\alpha_7 & -\alpha_1-\alpha_2\alpha_5\alpha_6\alpha_7\alpha_8 & -\alpha _2 \alpha _7 \\
 0 & 1 & 0 & -(\alpha_2+\alpha_3)\alpha_5\alpha_6 & -(\alpha_2+\alpha_3)\alpha_5\alpha_6\alpha_8 & -\alpha _2 \\
 0 & 0 & 1 & -(\alpha _4+\alpha _5) & -\alpha _5 \alpha _8 & 0 \\
\end{array}
\right)\,.\label{Cmat2}
\end{equation}
This $C$-matrix does not have any obvious positivity properties, ie. there is no choice of signs for edge variables $\alpha_i$ such that all main minors are positive. 

The lack of positive coordinates does not prevent us from calculating the on-shell functions for the BCFW term \eqref{fig:NonPlanarOnShell2} 
using the dual formulation (\ref{dual}) for the on-shell diagram \eqref{fig:NonPlanarOnShell2}, which gives the same result as direct evaluation of the product of two MHV amplitudes with $1/P^2$ pole,
\begin{equation}
    {\cal F} = \frac{A_4(3,4,\hat{5},I)A_4(I,6,\hat{1},2)}{s_{345}}\,,
\end{equation}
where $\lambda_{\hat{5}}$ and $\widetilde{\lambda}_{\hat{1}}$ being the shifted momenta and $I$ the internal on-shell leg. 


Unlike the planar case, where the cells have a simple combinatorial characterization and are understood to be associated with stratification of the positive Grassmannian, 
much remains to be understood about the connection between between non planar on-shell diagrams and the Grassmannian in general. In what follows, we find the Grassmannian geometry for non-adjacent BCFW terms. In other words, we identify a particular subspace in $G(k,n)$ associated with the non-planar on-shell diagrams which appear in the context of non-adjacent BCFW recursion relations. This is a generalization of the connection between on-shell diagrams/plabic graphs and the cells of the positive Grassmannian $G_+(k,n)$ into the non-planar sector, at least for the special set of BCFW cells identified. We start with a review of the MHV case which was worked out in \cite{Arkani-Hamed:2014bca} before discussing the new results on the NMHV non-planar cells. 

\subsection{MHV amplitudes}

It was shown in \cite{Arkani-Hamed:2014bca} that any non-planar MHV on-shell diagram evaluates to a linear combination of Parke-Taylor factors with coefficients $+1$. For instance: 
\input{figures/MHVgeneralconfig}

\vspace{1.2cm}
This statement follows from the fact that the whole $G(2,n)$ Grassmannian can be decomposed into positive Grassmannians $G_+(2,n)$ with various orderings, and holds for arbitrary complicated MHV diagrams which are associated with a cell in $G(2,n)$.

This is obvious from the geometric picture. The real Grassmannian $G(2,n)$ can be represented as a collection of $n$ points on the projective line $\mathbb{P}^1$ (with no restrictions). Obviously, any configuration of these points has a certain ordering, so any point in $G(2,n)$ is also in one of the positive Grassmannians $G_+(2,n)$ (with a particular ordering). An arbitrary non-planar on-shell diagram then corresponds to a union of $G_+(2,n)$ with different orderings.

The non-adjacent BCFW recursion relations produce a special class of non-planar on-shell diagrams. We expect this class to be particularly simple because in the recursion we have only one type of factorization diagram into $(n{-}1)$-point MHV amplitude and $3$-point $\overline{\rm MHV}$ amplitude (the latter necessarily involve leg $3$ instead of $1$).

Let us start our discussion with the $(31)$ shift. At 4-point there are two contributing terms:
\input{figures/BCFW4pt}
Following the discussion in Section 2, both factorization diagrams are inverse soft factors on the 3-point amplitude $A(124)$. The first diagram then evaluates to $-A(1324)$ and the second to $-A(1342)$, and the underlying Grassmannian geometry are just two different configurations of points on the projective line. So in this case each diagram is one particular fixed ordering of points on a line in $\mathbb{P}^1$. 
\input{figures/MHV4pt}
At 5-point we get again two factorization diagrams, 
\input{figures/BCFW5pt}
The first diagram adds points $3$ between $1$ and $2$ in an ordered amplitude $A(1245)$ and leads to $-A(13245)$, while for the evaluation of the second diagram we need to use the representation (\ref{fig:MHV4pt}), and we get $-A(13425) - A(13452)$. Note that in the first diagram $\hat{1}$ and $I$ are adjacent while in the second diagram they are non-adjacent. The formula for the amplitude is then
\begin{equation}
    A_{5,2} = -A(13245) - A(13425) - A(13452)\,.
\end{equation}
This easily generalizes to the $n$-pt case, the BCFW recursion using $(31)$ shift gives,
\begin{equation}
    A_{n,2} = -\sum_{j=3}^{n} A(134{\dots}j\,2\,j{+}1{\dots}n)\,.
\end{equation}

This is just a formal way to write that we start with the $(n{-}1)$ labels $1,3,4,5,\dots,n$ and we insert label $2$ anywhere between $3,\dots,n$. In order to show that this equal to the original Parke-Taylor factor $A(123\dots n)$ we use a U(1) decoupling identity,
\begin{equation}
    \sum_{j} A(13\dots j\,2\,j{+}1\dots n) = 0\,,
\end{equation}
where we sum over all possible positions of the label $2$. 

It is easy to see that for the general $(k1)$ shift the BCFW recursion relations contain two terms,
\input{figures/BCFWMHV}
Each term is one particular on-shell diagram 
\input{figures/Onshell2}
and they evaluate to
%
\begin{equation}
 \label{MHVbcfw}
    t_1 = (-1)^k \sum_{\sigma\in \Sigma_1} A(1\,k\,k{-}1 \sigma),\qquad 
    t_2 = (-1)^k \sum_{\sigma\in \Sigma_2} A(1\,k\,k{-}1 \sigma)\,,
\end{equation}
where $\Sigma_1=\{2,\cdots,k{-}2\}^T\shuffle\{k{+}1,\cdots,n\}$ and $\Sigma_2=\{2,\cdots,k{-}1\}^T\shuffle\{k{+}2,\cdots,n\}$. Here $\shuffle$ denotes the shuffle product of two orderings while the superscript $^T$ denotes the reversal of an ordering.
The corresponding geometry is just given by a union of $G_+(2,n)$s with a particular ordering given by the ordering of the Parke-Taylor factor in the sum. For the amplitude we get
\begin{equation}
    A_{n,2} = -\sum_\sigma A(1\,k\,\sigma_{2{\dots} k{-}1, k{+}1{\dots}n})\,,
\end{equation}
where we sum over all Parke-Taylor factors with indices $1,k$ being adjacent and over all permutations of other labels which keep labels $k{-}1,{\dots},2$, resp. $k{+}1,{\dots},n$ relatively ordered. The fact that this is equal to a single Parke-Taylor factor with canonical ordering $A(123\dots n)$, is the famous Kleiss-Kuijf (KK) relation \cite{Kleiss:1988ne}. 

Note that the on-shell diagrams we got in the context of BCFW recursion relations formed a subset of all diagrams, and only special labels appeared in the linear combinations of Parke-Taylor factors (\ref{MHVbcfw}). However, as noted earlier any on-shell diagram evaluates to some combination of Parke-Taylor factors and the underlying geometry is just a union of ordered points on a projective line. The first case where non-adjacent BCFW shifts lead to something qualitatively new is at the NMHV level.

\subsection{NMHV amplitudes}

For $k>2$ the non-planar on-shell diagrams are completely new on-shell functions, which can not be obtained by relabeling of planar on-shell functions, ${\cal R}$-invariants and their higher $k$ generalizations, or any linear combinations of them. To see what is new, we can look at an 8-dimensional non-planar on-shell diagram of $G(3,6)$ 
\begin{equation}
 \includegraphics[scale=0.9, valign=c]{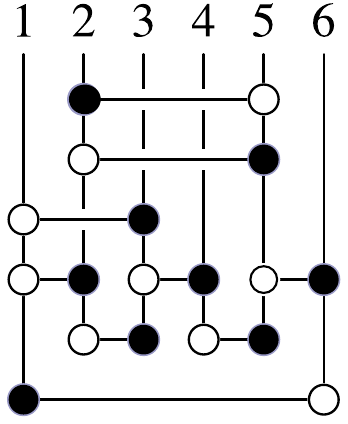}  = \frac{\langle 12\rangle [64]\ \delta^{3\times 4}(C^*\cdot \tilde \eta) \  \delta^{2\times 2}(\lambda\cdot\tilde\lambda)}{\langle 13\rangle[56]\langle 1|5{+}6|4]\langle 2|4{+}5|6](\langle 23\rangle[56]\langle 1|5{+}6|4] {-} \langle 12\rangle[45]\langle 3|4{+}5|6\rangle)}.
\end{equation}
Note the pole $\la23\ra[56]\la1|56|4] - \la12\ra[45]\la3|45|6]$ does not factorize, and can not be simplified. This is a new type of singularity of purely origin that would never show up in the planar on-shell diagrams. It also implies that the non-planar loop amplitudes have new kinematical poles (in external momenta) which do not arise in the planar limit. 

This on-shell diagram is one of all the inequivalent 8-dimensional cells of $G(3,6)$ (i.e. leading singularities) identified in \cite{Bourjaily:2016mnp}, where new type of on-shell functions with more complicated poles were found. The authors of \cite{Bourjaily:2016mnp} also classified all inequivalent 9-dimensional top cells of $G(3,6)$, but their underlying Grassmannian geometry (if any) is still unclear.

In this paper we initiate the exploration of non-planar Grassmannian geometry by looking at the BCFW on-shell diagrams (we also refer to them as BCFW cells) which are expected to form a particular simple subset of all non-planar on-shell diagrams. Our goal is to identify the Grassmannian geometry for these special cells. Specifically, for each diagram parameterized by some $G(3,n)$ matrix, we want to  associate a (generally non-convex) configuration of $n$ points in $\mathbb{P}^2$ to it. Once identified, this configuration sets the signs of all minors needed to define the subspace in $G(3,n)$ corresponding to the on-shell diagram.

\subsubsection*{Five point amplitude}

Let us start with the simplest 5-point case. Note that for the adjacent BCFW shift, say $(51)$, we only had one BCFW diagram which was interpreted geometrically as 
\input{figures/BCFWgeom}
This is nothing else than the convex configuration of five points, ie. top cell of $G_+(3,5)$. The three lines we chose to draw are arbitrary as no three points are on a single line. 

For the non-adjacent BCFW shift $(41)$ we get a sum of two terms, 
\input{figures/BCFWonshell}
Let us look at the first term, by definition this is equal to
\begin{equation}
    \frac{A^{k=2}(3,\hat{4},5,I)A^{k=2}(I,\hat{1},2)}{s_{12}}\,.
\end{equation}
Using the procedure from the previous section, the first MHV amplitude $A^{k=2}(3,\hat{4},5,I)$ corresponds to a line with $3,\hat{4},5,I$ (in this ordering), the second MHV amplitude gives a line $I,1,2$, and the point $4$ is on the line $1,\hat{4}$. Similar analysis can be done for the second BCFW term, and we get two configurations 

\begin{equation}
    \input{figures/Non5pt_config2_a} \quad \quad \hspace{1cm} \input{figures/Non5pt_config2_b} \label{Non5pt_config2}
\end{equation}
Note that points $I$, $\hat{4}$ are not part of the original set of points and do not participate in the positivity/negativity constraints on the minors of the $C$-matrix.

\subsubsection*{Grassmannian cells and positivity conditions}

Now we want to identify the cells in $G(2,5)$ which correspond to the two configurations (\ref{Non5pt_config2}). The configuration on the right is simple, it is just a usual positive Grassmannian top cell with an ordering $1,5,2,3,4$, and hence satisfies $(ijk)>0$ where $i,j,k$ respect this ordering.  
\begin{equation}
    \input{figures/MinorSigns2} \quad \quad \begin{aligned} 
    &(152)>0 \quad (123)>0 \quad  (523)>0 \quad (234)>0 \\
    &(153)>0 \quad (124)>0 \quad (524)>0 \\
    &(154)>0 \quad (134)>0  \quad (534)>0
    \end{aligned}
\end{equation}
The configuration on the left is more interesting, and it is a genuinely new non-planar $G(2,5)$ cell. The points $1,2,3,4$ form a convex configuration and hence $(123)$, $(124)$, $(134)$, $(234)>0$. If point $5$ was next to $3$ (say to the left), we would have $(125)$, $(135)$, $(235)$, $(154)$, $(354)$, $(254)>0$ and this would be still just a top cell of $G_+(3,5)$ with ordering $1,2,3,5,4$. However, moving the point $5$ away from $3$ to a different segment of the line changes the signs:
%
\begin{equation}
    \input{figures/MinorSigns1} \quad \quad \begin{aligned} 
    &(123)>0 \quad (234)>0 \quad  (235)>0 \quad {\bf (254)\lessgtr 0}\\
    &(124)>0 \quad (125)>0 \quad {\bf (154)<0} \\
    &(134)>0 \quad (135)>0  \quad (354)>0
    \end{aligned} \label{non5pt1}
\end{equation}
We see that two signs flipped: point 5 is now to the left of the line $(41)$, hence $(415)=(154)<0$, and 5 can be on either side of the line $(24)$ depending on the exact positions of 2,4, hence $(254)$ can have any sign and it does not impose any condition. Here we should remind the reader that each column of the $C$-matrix can be always rescaled without changing the Grassmannian cell. Hence, we can flip $(j)\rightarrow -(j)$ for any column and the configuration of points is still the same. Therefore, in (\ref{non5pt1}) we can flip $(5)\rightarrow -(5)$ and get
\begin{equation}
   \begin{aligned} 
    &(123)>0 \quad (234)>0 \quad  {\bf (235)<0} \quad {\bf (254)\lessgtr 0}\\
    &(124)>0 \quad (125)>0 \quad (154)>0 \\
    &(134)>0 \quad (135)>0  \quad {\bf (354)<0}
    \end{aligned}
\end{equation}
and it should correspond to the same configuration. It is not hard to see that these signs of minors are reproduced by 
\begin{equation}
    \input{figures/Non5pt_config2_a_alternative} \label{non5pt2}
\end{equation}
which is indeed the same configuration when we wrap $5$ around infinity and come back on the right segment. As a result, drawing a point on the left or right outer segments are equivalent as (\ref{non5pt1}) and (\ref{non5pt2}) are the same configurations corresponding to the same non-planar cells in $G(2,5)$.

At 5-point, there are really only two inequivalent BCFW shifts: adjacent $(51)$ and next-to-adjacent $(41)$; all others are related by cyclicity. But at higher points, we have a general $(k1)$ shift. Below, we first study the next-to-adjacent shift $(n{-}1\,1)$ before considering a general case.

\subsubsection*{Six-Point Amplitude}

Let us consider the calculation of the 6-pt NMHV amplitude using $(51)$ BCFW shift. We get four terms where the amplitude factorizes into two MHV amplitudes and two other terms where it factorize into a 5-pt NMHV and a 3-pt $\overline{\rm MHV}$ amplitudes. 
\input{figures/6ptBCFW}
As before, we can easily express these BCFW terms as non-planar on-shell diagrams. The first term, for instance, is given by a single on-shell diagram:  
%
\input{figures/Onshell_6pt_nonplanar_1.tex}
%
and similarly for the three other terms  in \eqref{eq:6pt-bcfw-15} of this type. They are are just products of two MHV amplitudes with shifted labels, and we can construct the Grassmannian geometry in the same way as for the 5-pt amplitude, 
\input{./figures/Non6pt_config.tex}

The last two terms of the form $\mathcal A_3^{(1)}\otimes \mathcal A_5^{(3)}$ have the interpretation of (holomorphic) inverse soft-factors, and the associated configurations can be obtained by adding point $5$ to some fixed configurations of points $1,2,3,4,6$. Now the resulting geometry does depend on how we recurse the 5pt NMHV amplitude, whether two adjacent or non-adjacent legs in the sub-amplitude are chosen for BCFW shift. To get a closed form formula, which generalizes for any $n$ and any $(k1)$ shift, we use the $(I1)$ shift, ie. the legs which are shifted in the 6-pt recursion are also used to represent the 5-pt amplitude. 
For the bottom left term in \eqref{eq:6pt-bcfw-15}, $(I1)$ shift is an adjacent shift, and we start with the configuration for the top cell of $G_+(3,5)$ with labels $1,2,3,4,6$ and add point $5$ between $1$ and $6$. We got a positive Grassmannian cell in $G_+(3,6)$ with minor $(156)=0$ while all others being positive. 
\begin{equation}
    \input{figures/Gr35pointadd}
\end{equation}
For the last term of \eqref{eq:6pt-bcfw-15}, the $(I1)$ shift is a non-adjacent shift with which we get a representation of the 5pt NMHV amplitude $A_5^{k=3}(1,2,3,4,6)$ as in \eqref{eq:non-planar-bcfw-5pt} as a sum of two on-shell diagrams 
\begin{equation}
    \input{figures/6ptNMHVonshell}\label{onshell6}
\end{equation}
each of them is associated with one 8-dimensional cell in $G(3,6)$. Adding point $5$ between $4$ and $1$, we get a sum of two configurations 
\begin{equation}
    \input{figures/Non6pt_config_2} \label{fig:Non6pt_config_2}
\end{equation} 

As before, we can easily read off the signs of all $(3\times 3)$ minors and hence fully determine the 8-dimensional cell in $G(3,6)$. For the first configuration in \eqref{fig:Non6pt_config_2} it is 
\begin{equation}
    \input{figures/Non6pt_config2_a} \quad \begin{aligned} 
    &(123)>0 \quad (134)>0 \quad  (146)<0 \quad (236)<0 \quad (345)>0\\
    &(124)>0 \quad (135)>0 \quad (156)<0 \quad (245)>0\quad (346)>0\\
    &(125)>0 \quad (136)<0  \quad (234)>0 \quad (246)\lessgtr0\quad (356)>0\\
    & (126)<0 \quad (145)=0 \quad (235)>0\quad (256)\lessgtr0\quad (456)>0\\
    \end{aligned} \label{non6pt1}
\end{equation}
The second diagram is just a usual cell in $G_+(3,6)$ with an ordering $1,6,2,3,4,5$ and $(145)=0$, as also suggested by the right on-shell diagram in (\ref{onshell6}). However, as we will see later we can also interpret this term as a configuration 
\begin{equation}
    \input{figures/Non6pt_config2_b} \quad \quad \begin{aligned} 
    &(123)>0 \quad (134)>0 \quad  (146)<0 \quad (236)<0 \quad (345)>0\\
    &(124)>0 \quad (135)>0 \quad (156)<0 \quad (245)>0\quad (346)\lessgtr0\\
    &(125)>0 \quad (136)<0  \quad (234)>0 \quad (246)\lessgtr0\quad (356)\lessgtr0\\
    & (126)<0 \quad (145)=0 \quad (235)>0\quad (256)\lessgtr0\quad (456)>0\\
    \end{aligned} \label{non6pt2}
\end{equation}
This means that the planar on-shell diagram on the right of (\ref{onshell6}) can be associated either with the cell in $G_+(3,6)$ or to the cell (\ref{non6pt2}) in $G(3,6)$. In both cases, we get the same canonical form (up to a sign). The union of these two cells in $G(3,6)$ is a space with arbitrary signs of any $(ij6)\lessgtr0$ (ie. point $6$ can be anywhere on that line) and vanishing form.

\subsubsection*{All-$n$ expansion}

We can extend this procedure to higher points using the next-to-adjacent $(n{-}1\,1)$ shift. The amplitude either factorizes into a product of two multi-particle MHV amplitudes or a product between an $(n-1)$-pt NMHV amplitude and a $3$-pt $\overline{\rm MHV}$ amplitude
\input{figures/BCFW_NMHV_gen.tex} 
Each BCFW term of the form MHV$\times$MHV, as in the adjacent case, is given by one on-shell diagram associated with a particular cell of $G(3,n)$, describing a single configuration of points where all except point $n{-}1$ are ordered on two lines (with different orderings) and the point $n{-}1$ is on another line. By contrast, every BCFW term of the form ($n{-}1$-pt NMHV) $\times$ (3-pt $\overline{\rm MHV}$) gives rise to a sum of on-shell diagrams, and correspondingly, a collection of configurations of $n$ points on three lines; all of these terms nicely complement the first type of configurations. Overall the $n$-pt NMHV amplitude is a sum over all configurations where points $1,2,3,\dots,n{-}1$ are ordered on three lines, exactly as in the $(n{-}1)$-pt NMHV amplitude using adjacent shift $(n{-}1\,1)$, but in addition we insert point $n$ on one of the three lines at an arbitrary position -- denoted by a circle:
\begin{equation}
    A_n^{k=3} = \sum_{\sigma_n} \sum_{i,j} \input{./figures/NMHVform1.tex} \label{NMHVnext}
\end{equation}
Note that points $1$ and $n-1$ are adjacent on the third line in all the configurations, and points $i,j$ marking the borders are such that there are at least two points on each line. Each configuration corresponds to an generally non-planar on-shell diagram, and the particular cell in $G(3,n)$ associated with it is defined by a set of inequalities which can be read off directly from the configuration. For illustration, we show below all $10$ configurations for the $7$-pt NMHV amplitude using $(61)$ shift: 
\input{figures/NMHV_7pt_61shift_all_configs.tex}
For the general shift $(k1)$, we proceed in the same way and find the following expression
\begin{equation}
    A_{n}^{k=3} = \sum_{\substack{i,j,l,m\\ \sigma(j+1\cdots k-1; m+1\cdots n)}} \input{./figures/NMHVform2.tex}\label{generalNMHV}
\end{equation}
Here again we have $n$ points on three lines distributed as follows: points $1$ and $k$ are fixed in adjacent positions on the third line; all other points are grouped into two ordered sets $S_1=\{2,3,\dots,k{-}1\}$ and $S_2=\{k{+}1,\dots,n\}$, and distributed among three lines. Schematically we denote the internal orderings as
\input{figures/NMHVform3.tex}
where blue represents $S_1$ while red represents $S_2$. In fact, because of projectivity, the picture is invariant under switching blue and red points. The set $S_1$ further splits into three pieces: $\color{blue}\{2,{\dots},i\}$ on the first line, $\color{blue} \{i{+}1,{\dots},j\}$ on the second line and $\color{blue} \{j{+}1,{\dots},k{-}1\}$ on the third line. Similarly the second set $S_2$ splits into $\color{red} \{k{+}1,{\dots},l\}$ on the first line (to the right), $\color{red}\{l{+}1,\dots,m\}$ on the second line (horizontal) and $\color{red}\{m{+}1,{\dots},n\}$ on the third line (to the left). On the first two lines the two sets do not interact -- they belong to distinct segments, and fixing the splitting points $i,j$, resp. $l,m$ determines the configuration on these two lines completely. On the third line, points from $S_1$ and $S_2$ both sets are mixed together while keeping the relative orderings within their respective subsets $\color{red}\{j{+}1,{\dots},k{-}1\}$ and $\color{blue}\{m{+}1,{\dots},n\}$. This means that even after fixing $i,j,l,m$ we get multiple configurations from various ordering of points on the third line. In (\ref{generalNMHV}) we sum over all the possible configurations.

To demonstrate how the points are distributed on the third line, consider an example for a general $(k1)$ shift where we have points $\color{blue} k{-}2,k{-}1$ from $S_1$ and points $\color{red} n{-}1,n$ from $S_2$ to be distributed on that line. This gives us following configurations on the third line,
\begin{equation}
    \input{figures/points_mixing_on_L3}
\end{equation}
To get a full geometry we decorate them with a (fixed) configuration of points on the first two lines. As an explicit example, consider the 8-pt configurations for the $(51)$ shift. For some choices of $i,j,l,m$ we end up with only one configuration for the third line
\input{./figures/Examples1.tex}
but for another choice $i=1$, $j=2$, $l=6$, $m=7$ we get three different terms, 
\input{./figures/Examples2.tex}
To get the amplitude we have to sum over all possible choices of $i,j,l,m$ and subsequent distributions of the remaining points on the third line. Note that for the adjacent shift $(n1)$, $S_2$ is empty and we get a usual representation \eqref{BCFWNMHV2}, while for the next-to-adjacent shift $(n{-}1\,1)$, $S_2=\{n\}$ and we reproduce the formula (\ref{NMHVnext}). We can also explicitly identify the cells in $G(3,8)$ by fixing signs of minors. For example, for the first figure in (\ref{8ptex1}) we get 
\begin{equation}
{\small  \begin{aligned} 
    &(123)>0 \quad (134)>0 \quad  (146)<0 \quad (167)<0 \quad (237)<0 \quad (256)<0\quad (347)=0 \quad (457)>0\\
    &(124)>0 \quad (135)>0 \quad (147)<0 \quad (168)>0\quad (238)>0\quad (258)<0\quad (348)>0\quad (458)<0\\
    &(125)>0 \quad (136)<0  \quad (148)>0 \quad (178)>0 \quad (245)>0 \quad (267)<0\quad (357)>0\quad (467)<0\\
    &(126)=0 \quad (137)<0 \quad (156)<0\quad (234)>0\quad (246)<0 \quad (268)>0\quad (358)<0 \quad (478)<0\\
    &(127)<0 \quad (138)>0 \quad (157)<0 \quad (235)>0 \quad (247)<0 \quad (345)>0\quad (367)<0\quad (567)<0\\
    &(128)>0 \quad (145)>0 \quad (158)=0 \quad (236)<0 \quad (248)>0 \quad (346)<0\quad (378)<0 \quad (568)>0\\
    & (578)>0\\
    \end{aligned} }
\end{equation}
where we skipped the minors which signs are not completely fixed. As before, we can wrap points 6 and/or 7 around infinity which flips signs of all minors with 6 and/or 7, but the cell in $G(3,8)$ remains unchanged.

In summary, all non-planar on-shell diagrams arising from non-adjacent BCFW recursion relations for NMHV amplitudes correspond to cells in the Grassmannian $G(3,n)$ and describe configurations of $n$ points located on three lines. For the adjacent shift $(n1)$ these configurations are all convex, corresponding to $(2n{-}4)$-dimensional cells in $G_+(3,n)$, while for a general shift the configurations subsume two independent orderings $S_1$, $S_2$. As expected, these geometries, despite being non-planar (or non-convex), are simpler than the general case of on-shell diagrams and cells in $G(3,n)$. We have seen earlier in \eqref{eq:planar-LS-conf-gen} that, even in the planar case, the geometry of a general $(2n{-}4)$-dimensional cell involve $n$ points on \emph{five} different lines in $\mathbb{P}^2$, but here all points are restricted to be on \emph{three} lines. It would be very interesting to find a compact combinatorial description of these non-planar cells. Unlike the planar case, where a single permutation $\pi(1,2,{\dots},n)$ specified the cell completely, the above discussion suggests that there should be at least two different permutations. 

\section{Non-planar tree-level \texorpdfstring{$\mathcal{R}$}{R}-invariants}
\label{sec:nonplanarR}

In the last section, we found a correspondence between the terms in non-adjacent BCFW recursion relation and Grassmannian geometries for NMHV amplitudes, ie. configurations of $n$ points in $\mathbb{P}^2$. We showed that these $n$ points are localized on three lines:
\input{./figures/General1.tex}

We refer to these $(2n{-}4)$-dimensional subspaces in the Grassmannian $G(3,n)$ as \emph{non-planar BCFW cells}. This is the most general configuration we would get in any BCFW recursion relation, for any shift and any choice of representation of the lower-point NMHV amplitudes in (\ref{BCFWNMHV4}). As we showed in the previous section, making consistent choices for shifts gives us a particularly simple and compact representation of the amplitude (\ref{generalNMHV}) which is a direct generalization of the expansion in terms of cells in $G_+(3,n)$ with the adjacent shift (\ref{BCFWNMHV2}). Nonetheless even if we make arbitrary choices for the lower-point geometries, we never get anything more complicated than (\ref{generalconf}), only that labeling the collection of terms that enter the BCFW formula for the NMHV amplitude is more complicated.

In this section, we discuss the superfunctions associated with these configurations. In other words, we are to find an analogue of ${\cal R}_{1,i{+}1,j{+}1}$ (\ref{R2}), the ${\cal R}$-covariant which is associated with the general planar BCFW configuration (\ref{ThreeLines2}). This is a laborious but in principle straightforward exercise. These configurations arise in the analysis of individual terms in the BCFW recursion relations, and each of them is tied to a particular non-planar on-shell diagram, as we showed in Section \ref{sec:nonadjBCFW}. We can compute the diagram either using (\ref{dual}) or by directly taking the product of 3-point amplitudes with a proper Jacobian to get the superfunction ${\cal F}_\gamma(\lambda,\widetilde{\lambda},\widetilde{\eta})$. For instance, for the diagram (\ref{fig:NonPlanarOnShell2}) we studied earlier, the Grassmannian configuration is 
%
\input{figures/NonPlanarOnShell}
and plugging the $C$-matrix (\ref{Cmat2}) obtained from boundary measurement into (\ref{dual}) we get
\begin{equation}
{\cal F}_\gamma = \frac{\Delta\cdot\la1|26|5]}{s_{345}\la 12\ra\la 16\ra[34][45]\la 2|34|5]\la 6|34|5]\la 1|26|3]}\,,\label{F1}
\end{equation}
where the super $\delta$-functions are
\begin{equation}
    \Delta = \delta^4(P)\,\delta^4(Q)\,\delta(\widetilde{\Xi})\quad\mbox{with}\quad \widetilde{\Xi} = [45]\widetilde{\eta}_3 + [53]\widetilde{\eta}_4 + [34]\widetilde{\eta}_5\,.
\end{equation}
We make two observations here. First, the kinematical part of (\ref{F1}) looks different from 6-point ${\cal R}$-invariants, it has more poles and a numerator factor. On the other hand, the fermionic $\delta$-function is identical to the planar case (up to relabeling). This is true in general for a simple but simple reason: the argument of the fermionic $\delta$-function $\delta^4(\Xi)$ in (\ref{Xi}) (or $\delta^4(\widetilde{\Xi})$ in the special boundary case) is agnostic of the orderings of points within a line, only the partition into three sets $P_1$, $P_2$, $P_3$ matters. 

For the planar R-invariant ${\cal R}_{1,i{+}1,j{+}1}$ (\ref{Xi}), we used $\Xi_{1,i{+}1,j{+}1}$ to denote this argument, which could have also been represented as 
\begin{equation}
\input{figures/NMHV_8pt_nonplanar_config.tex}\quad    \Xi_{1,P_2,P_3} \equiv \sum_{k\in P_2} \la k|P_2P_3|1\ra \widetilde{\eta}_k + \sum_{j\in P_3} P_2^2 \la 1j\ra \widetilde{\eta}_j\,, \label{Xi3}
\end{equation}
where we list $P_2$, $P_3$ explicitly ( $P_1$ is just a complement). For the non-planar generalization, indicating the boundary points is insufficient, and we need to use the more comprehensive symbol in order to specify the sets $P_2$, $P_3$.

\subsection{Non-planar \texorpdfstring{${\cal R}$}{R}-invariant}

In analogy with the definition of the ${\cal R}$-invariant ${\cal R}_{1,i{+}1,j{+}1}$ associated with the convex configuration (\ref{ThreeLines2}), we define the \emph{non-planar ${\cal R}$-invariant} associated with a general non-convex configuration arising from the non-adjacent BCFW recursion,
\input{./figures/General2.tex}\label{genconf}
In general, the points are completely unordered and we do need to specify all five subsets $P_1^a$, $P_1^b$, $P_2^a$, $P_2^b$, $P_3$ including their internal orderings. While the non-planar ${\cal R}$-invariant can be a very complicated kinematical function, the actual form is surprisingly simpler. The $\delta$-functions are given by
\begin{equation}
    \Delta_{1,P_2,P_3} = \delta^4(P)\,\delta^8(Q)\,\delta^4(\Xi_{1,P_2,P_3})\,,
\end{equation}
where $\Xi_{1,P_2,P_3}$ was defined in (\ref{Xi3}). Here we denote
\begin{equation}
    P_1 = P_1^a  \cup P_1^b,\qquad P_2 = P_2^a  \cup P_2^b\,.
\end{equation}
As noted above, the super $\delta$-function neither depends on the ordering of points, nor on the split of $P_2$ into $P_2^a,P_2^b$, and is analogous to $\Delta_{1,i{+}1,j{+}1}$ which appeared in the ${\cal R}$-invariant. 

More surprisingly, the bosonic function takes exactly the same form as its planar counterpart (\ref{R2}), given by the product of three Parke-Taylor factors
\begin{align}
    {\cal R}_{1,\{P_1^a,P_1^b\},\{P_2^a,P_2^b\},P_3} &= PT(1,P_1^a,I_1,P_1^b)\times PT(I_2,P_2^a,I_1,P_2^b)\times PT(I_2,P_3,1) \nonumber\\
    & \qquad \times \la 1|P_2P_3|1\ra^3\cdot \Delta_{1,P_2,P_3}\,. \label{Rinv4}
\end{align}
The $\lambda$ spinors for $I_1$, $I_2$ are given by the same formulae (\ref{spinors}),
\begin{equation}
    \lambda_{I_1} = \la 1|P_3P_2,\qquad \lambda_{I_2} = \la 1|P_1P_2\,.
\end{equation}
The only difference between (\ref{Rinv4}) and (\ref{R2}) is that points $1,I_1$, and $I_1,I_2$ are not adjacent and that leads to more types of poles that involve $I_1$ and $I_2$. Also, in the planar case (\ref{R2}) the factor $\la 1|P_2P_3|1\ra^3$ always canceled against poles from the Parke-Taylor factors and ${\cal R}_{1,i{+}1,j{+}1}$ never had any numerator factors apart from delta functions. In the non-planar case, this is no longer true, and we get up to two factors of $\la 1|P_2P_3|1\ra$ left in the numerator.

\subsection*{Examples}

First, let us reproduce the expression (\ref{F1}) we obtained earlier from the direct computation using on-shell diagram, $C$-matrix and (\ref{dual}). We draw the same configuration again and label points $I_1$, $I_2$,
\input{figures/NonPlanarOnshellwithI}
Their $\lambda$-spinors are given by
\begin{equation}
    \lambda_{I_1} = \la 1|(5)(34) = [5|(34)\cdot \la 15\ra, \qquad \lambda_{I_2} = \la 1|(26)(34)\,.
\end{equation}
We see that the $\lambda_{I_1}$ spinor simplifies because there is only one point on the third line, but we keep the constant factor $\la 15\ra$ in the definition of $\lambda_{I_1}$ -- we will see that it trivially cancels against the same factors in the numerator. The argument of the fermionic delta function is given by
\begin{align}
    \Xi_{1,34,5} = &\la 3|(34)(5)|1\ra\widetilde{\eta}_3 + \la 4|(34)(5)|1\ra\widetilde{\eta}_4 + s_{34}\la 15\ra\widetilde{\eta}_5\nonumber\\
    = &\la15\ra\la34\ra ([45]\widetilde{\eta}_3 + [53]\widetilde{\eta}_4 + [34]\widetilde{\eta}_5)\,.
\end{align}
The Parke-Taylor factors evaluate to
\begin{align}
    PT(1,2,I_1,6) &= \frac{1}{\la12\ra\la 2I_1\ra\la I_16\ra\la 61\ra} = \frac{1}{\la12\ra\la16\ra\la 2|34|5]\la6|34|5]\la16\ra\cdot\la15\ra^2}\,,\\
    PT(I_1,3,4,I_2) &= \frac{1}{\la I_13\ra\la34\ra\la 4I_2\ra\la I_2I_1\ra} = \frac{1}{[34][45]\la1|26|3]\la1|26|5]\cdot \la15\ra^2 \la34\ra^3}\,,\\
    PT(1,5,I_2)&= \frac{1}{\la15\ra\la 5I_2\ra\la I_21\ra} = \frac{1}{s_{345}\la 1|26|5]\la15\ra^3}\,,
\end{align}
where we used the momentum conservation $P_1+P_2+P_3+p_1=0$ to rewrite $\la 1|P_1P_3|1\ra = -\la 1|P_1P_2|1\ra = \la 1|P_3P_2|1\ra$ etc., and then $\la 1|P_1P_3|1\ra = \la 1|26|5]\la15\ra$. Plugging all these ingredients into (\ref{Rinv4}) we get 
\begin{equation}
    {\cal R}_{1,\{2,6\},\{34,\},\{5\}} = \frac{\la1|26|5]\cdot \delta^4(P)\,\delta^8(Q)\,\delta^4([45]\widetilde{\eta}_3 + [53]\widetilde{\eta}_4 + [34]\widetilde{\eta}_5)}{s_{345}\la 12\ra\la 16\ra[34][45]\la 2|34|5]\la 6|34|5]\la 1|26|3]}\,, \label{nonR6}
\end{equation}
which is equal to (\ref{F1}).

Let us look now at a more general 9-point configuration,
\input{./figures/Examples4.tex}
The internal spinors $I_1$, $I_2$, are now
\begin{equation}
    \lambda_{I_1} = \la 1|(89)(567),\qquad \lambda_{I_2} = \la1|(234)(567)\,,
\end{equation}
and following the same procedure we get
\begin{equation}
    {\cal R} = \frac{\la 1|(234)(89)|1\ra^2\cdot\delta^4(P)\,\delta^8(Q)\,\delta^4(\Xi_{1,567,89})}{\begin{array}{c}\la12\ra\la23\ra\la14\ra\la56\ra\la89\ra\la91\ra\la1|(89)(567)|3\ra\la1|(89)(567)|4\ra\la1|(89)(67)|5\ra\la1|(89)(56)|7\ra\\ \la 1|(234)(57)|6\ra\la1|(234)(56)|7\ra\la1|(234)(567)|8\ra\end{array}}\,,
\end{equation}
where we denoted ${\cal R} \equiv {\cal R}_{1,\{23,4\},\{56,7\},\{89\}}$. For a general configuration, we can plug into (\ref{Rinv4}) for the Parke-Taylor factors and obtain an expression for a general ${\cal R}$, 
\input{figures/NonPlanarRInv1}
where in the first term in the denominator we took the product of all $\la ab\ra$ of the points $a,b$ adjacent on one of the lines. There is a boundary case for which $P_2^b$ is empty, and the new pole $\la I_1\,I_2\ra$ appears from the Parke-Taylor factor.
\input{figures/NonPlanarRInv2}
Because of $\la I_1\,I_2\ra = \la 1|P_3P_1|1\ra P_2^2$ one power in the numerator cancels and we get only a single power of $\la 1|P_1P_3|1\ra$. If both $P_2^b$ and $P_1^b$ are empty, points $i_3,i_4,j_3$ are missing and $i_1\rightarrow i$, $i_2\rightarrow i{+}1$, $j_1\rightarrow j$, $j_2\rightarrow j{+}1$, the formula simplifies to 
\input{figures/NonPlanarRInv3}
which is just the usual ${\cal R}$-invariant (\ref{Rinv}). 

The general expression (\ref{Rinv5}) has a non-trivial numerator $\la 1|P_1P_3|1\ra^2$ (the third power cancels against the $\la 1P_2\ra$ pole), and poles from the Parke-Taylor factors. Note that 7 of these poles depend on points $I_1$, $I_2$ (these are non-holomorphic after plugging in expressions for $\lambda_{I_1}$, $\lambda_{I_2}$). This is in contrast with ${\cal R}_{1,i{+}1,j{+}1}$, where there is no extra numerator factor and only 5 non-holomorphic poles appear in the denominator (\ref{Rinv6}).

\subsection{Planar expansion}

 There is similarity between (\ref{Rinv5}) and the compact formula for MHV on-shell diagrams \cite{Arkani-Hamed:2014bca}, which also has a square of certain kinematical factors in the numerator. In that case, the formula was valid for all on-shell diagrams (of dimensionality $2n{-}4$), whereas our construction is only for BCFW cells. Nevertheless, we can explore the similarity further. As shown in \cite{Arkani-Hamed:2014bca} any MHV on-shell diagram (on-shell function) can be expressed as a linear combination of Parke-Taylor factors with various orderings, and $\pm1$ coefficients. We can ask the same question in the context of NMHV BCFW cells and formula (\ref{Rinv4}) -- can we express it as a linear combination of the original ${\cal R}$-invariants with different orderings? 

Let us first look at the case where points are ``misplaced" only on one of the lines while the rest are canonically ordered,
\input{./figures/SingleLine.tex}
The superfunction can be built from Parke-Taylor factors as
\begin{equation}
    {\cal R} = PT(1,{\dots},i,I_1)PT(I_1,i{+}1,{\dots},j,I_2,j{+}1,{\dots},k)PT(I_2,k{+}1,{\dots},n,1)\la 1|P_1P_3|1\ra^3\cdot \delta^4(\Xi) \label{Rexp}
\end{equation}
where $P_1=p_2+{\dots}+p_i$, $P_3=p_{k{+}1}+{\dots}+p_n$, and $\Xi=\Xi_{1,i{+}1,k{+}1}$, just as in the standard $R$-invariant. The non-planarity of the configuration is caused by the middle Parke-Taylor factor where points $I_1$, $I_2$ are not adjacent. We can use the Kleiss-Kuijf (KK) relations to rewrite this expression as a sum over Parke-Taylor factors where $I_1$ and $I_2$ are adjacent,
\begin{equation}
    PT(I_1,i{+}1,{\dots},j,I_2,j{+}1,{\dots},k) = \sum_\sigma PT(I_1,\sigma,I_2)\,,\label{KK1}
\end{equation}
where $\sigma$ belongs to the shuffle product of two sets
\begin{equation}
    \sigma \in \{i{+}1,\dots,j\}\shuffle \{j{+}1,\dots,k\}^T\,.
\end{equation}
This means we sum over permutations of labels $\{i{+}1,\dots,k\}$ where the relative orderings of both sets $\{j,j{-}1,\dots,i{+}1\}$ and $\{j{+}1,\dots,k\}$ (this is the transverse of the original set) are preserved. For example, in the case of $i=1$, $j=3$ and $k=5$ with $\sigma\in \{2,3\}\shuffle \{4,5\}^T$,
\input{figures/KKexample}
Note that while the KK relation is for MHV amplitudes, it also true for the Parke-Taylor factors as the super $\delta$-function is the same for any ordering. We can now plug (\ref{KK1}) into (\ref{Rexp}). Each term in the sum leads to a planar $R$-invariant with some given ordering. As a result, we get an expansion of ${\cal R}$ as a linear combination of $R$-invariants,
\input{./figures/SingleSum.tex}
%
Note that this is only possible because neither the Jacobian factor $\la 1|P_1P_3|1\ra$ nor the fermionic delta function $\delta^4(\Xi)$ depends on the ordering of points in $P_2$ and they are the same for all terms in the sum (\ref{Rexp2}).

As an example, let us expand the first of the 6-pt NMHV non-planar on-shell functions (\ref{fig:Non6pt_config}). In that case, the KK relation produces two terms,
\input{./figures/Example6pt.tex}
And we get two $R$-invariants for $(123)=0$ corresponding to orderings $1,2,3,4,5,6$ and $1,2,3,5,4,6$. The bosonic part of the on-shell functions (the fermionic part is the same across all term as stressed before) yield (\ref{nonR6}), 
\begin{align}
   & \frac{1}{s_{123}\la12\ra\la23\ra[45][56]\la1|23|4]\la3|45|6]} + 
    \frac{1}{s_{123}\la12\ra\la23\ra[45][46]\la1|23|5]\la3|45|6]}\nonumber\\
    &\hspace{5cm}= 
    \frac{\la 1|23|6]}{s_{123}\la12\ra\la23\ra[46][56]\la1|23|4]\la1|23|5]\la3|45|6]}\,.
\end{align}
Note that the pole $[45]$ is spurious, ie. it is present in both ${\cal R}$-invariants but cancels in the sum. Geometrically $[45]=0$ corresponds to merging points $I_1$ and $I_2$ and indeed that is not a singularity of the non-planar configuration. The appearance of the spurious poles (which cancel in the sum) is also a feature of the Parke-Taylor expansion of the on-shell functions for MHV on-shell diagrams \cite{Arkani-Hamed:2014bca}. Finally, we can do the same for the general case where points are misplaced on two lines. 
\input{./figures/DoubleLine.tex}
The expansion in terms of Parke-Taylor factors is then 
\begin{align}
    {\cal R} &= PT(1,{\dots},i,I_1,i{+}1,{\dots},j)\times PT(I_1,j{+}1,{\dots},k,I_2,k{+}1,\dots,m)\nonumber\\
    & \hspace{2cm}\times PT(I_2,m{+}1,\dots,n,1)\times \la 1|P_1P_3|1\ra^3\cdot \delta^4(\Xi_{1,j{+}1,m{+}1)}\,.\label{Rgen}
\end{align}
where $P_1=p_2+\dots+p_i+p_{i{+}1}+\dots+p_j$ is defined as usual, but there are now two sets of momenta on the interval between $2$ and $j$, and  $P_3=p_{m{+}1}+\dots+p_n+p_1$. The argument of the fermionic $\delta$-function is again the same as in a planar $R$-invariant for any ordering of points on three lines.

Now we use KK relations on both lines and express Parke-Taylor factors in terms of the ones where $1$, $I_1$, and $I_1$, $I_2$ are adjacent,
\begin{align}
   PT(1,2,{\dots},i,I_1,i{+}1,{\dots},j) &= \sum_{\sigma_1} PT(1,\sigma_1,I_1)\,,\\
   PT(I_1,j{+}1,{\dots},k,I_2,k{+}1,\dots,m) &= \sum_{\sigma_2} PT(I_1,\sigma_2,I_2)\,,
\end{align}
where the permutations $\sigma_1$, $\sigma_2$ are again given by shuffle products,
\begin{equation}
    \sigma_1 = \{2,{\dots},i\}\shuffle\{i{+}1,{\dots},j\}^T,\qquad
    \sigma_2 = \{j{+}1,{\dots},k\}\shuffle\{k{+}1,{\dots},m\}^T\,.
\end{equation}
We plug back into (\ref{Rgen}) and get 
\input{./figures/DoubleSum.tex}
This is a very interesting result. While a general non-planar ${\cal R}$-invariant is a new kinematical object associated with a non-convex configuration of points in $\mathbb{P}^2$, it can be expressed as a linear combination of ordinary ${\cal R}$-invariants with various orderings. This further shows that the BCFW cells form a very special subset of all leading singularities, ie. $2n{-}4$ dimensional on-shell diagrams.

\section{Kinematical dlog forms}
\label{sec:dlog}
In the previous section, we see how the Grassmannian configurations directly embody superfunctions of external kinematic data, sidestepping the standard procedures of computation from constructing representative $C$-matrices to evaluating contour integrals. In this process, the canonical forms associated with the BCFW cells appear as auxiliary objects to be integrate over with a set of $\delta$-functions. On the other hand, it was shown in \cite{He:2018okq} that super-amplitudes (or super-functions for individual BCFW cells) can be viewed more fundamentally as differential forms in the on-shell kinematic space. Operationally the kinematical differential forms can be obtained as a pushforward of the canonical forms on Grassmannian cells to the kinematic space. In this section, we discuss the connection between Grassmannian configurations and the kinematical forms with logarithmic singularities in ${\cal N}=4$ SYM. In particular, we provide a method for constructing a \emph{holomorphic dlog} representation of the kinematical forms for NMHV amplitudes, in both the planar and non-planar cases, without invoking the pushforward map. 



\subsection{Super-functions as kinematical differential forms}
Let us first review the idea of super-functions for individual BCFW cells as differential forms in the kinematical space. This was first formulated in the momentum twistor space \cite{Arkani-Hamed:2017vfh}. The relation between the dlog form $\Omega_Z$ and the superfunction is just a simple replacement
\begin{equation}
    {\cal F}(Z,\eta) = \Omega(dZ_k\rightarrow \eta_k)\,,
\end{equation}
where $\eta$ are momentum twistor Grassmann variables. Summing forms on individual cells gives the tree-level amplitudes (and loop integrands) as forms with logarithmic singularities on the boundaries of the Amplituhedron space.

It was suggested in \cite{He:2018okq} how this picture extends to the spinor-helicity space and this was used in the formulation of the momentum Amplituhedron \cite{Damgaard:2019ztj}.
We start with a super-function for a given on-shell diagram $\gamma$ in the non-chiral space, which is given by a Fourier transformation of ${\cal F}$ on half of the Grassmann variables $\widetilde{\eta}^I$,
\begin{equation}
   {\cal F}^{\gamma}_{n,k} = \oint \omega^{\gamma}_{n,k}\,\delta^{2k}(C\cdot \widetilde{\lambda})\delta^{2(n-k)}(C^\perp{\cdot}\lambda)\delta^{0|2k}(C\cdot \widetilde{\eta})\delta^{0|2(n-k)}(C^{\perp}\cdot \eta)\,. \label{Grassform2}
\end{equation}
In the case of BCFW cells the dimensionality of the form is $m=2n-4$. The non-chiral super function can be thought of as a differential form of degree $(2(n-k), 2k)$ in $(d\lambda, d\tilde\lambda)$ space with the replacement $\eta^{1,2}\rightarrow d\lambda^{1,2}$ and $\widetilde{\eta}^{1,2}\rightarrow d\widetilde{\lambda}^{1,2}$. The resulting $2n$-form vanishes identically as the super function contains $\delta^4(P)\delta^4(Q)\delta^4(\tilde Q)$. Upon the replacement
\begin{align}
    &\delta^4(Q) \to (dq)^4=\bigwedge_{\alpha=1}^2 \bigwedge_{\dot \alpha=1}^2 (dq)^{\alpha\dot{\alpha}} = \bigwedge_{\alpha=1}^2 \bigwedge_{\dot \alpha=1}^2 \left[\sum_{a=1}^n \lambda_a^\alpha(d\widetilde{\lambda}_a)^{\dot{\alpha}}\right]\,,\nonumber\\
    &\delta^4(\tilde Q) \to (d\tilde q)^4=\bigwedge_{\alpha=1}^2 \bigwedge_{\dot \alpha=1}^2 (d\tilde q)^{\alpha\dot{\alpha}} = \bigwedge_{\alpha=1}^2 \bigwedge_{\dot \alpha=1}^2 \left[\sum_{a=1}^n (d\lambda_a)^\alpha \widetilde{\lambda}_a^{\dot{\alpha}}\right]\,,
\end{align}
by virtue of momentum conservation 
\begin{equation}
(dq)^{\alpha\dot{\alpha}}  + (d\widetilde{q})^{\alpha\dot{\alpha}} = 0\,,    
\end{equation}
the full “super momentum-conserving form” vanishes. We can factor out half of the super momentum-conserving factor explicitly by partially localizing $C$ so that the first two rows are simply $(\lambda_1, \dots, \tilde \lambda_n)$.
As a result, we get 
\begin{equation}
     {\cal F}^{\gamma} = \delta^4(P) \delta^4(Q) \times \tilde {\cal F}^{\gamma}(\lambda,\tilde\lambda,\eta,\tilde\eta)\,. 
\end{equation}
Stripped of $\delta^4(P)\delta^4(Q)$, the remaining superfunction $\tilde {\cal F}_\gamma$ turns into a form,
\begin{align}
   \Omega_{n,k}^{\gamma} &= \int \omega_{n,k}^\gamma \prod_{\mu} \delta^2(C_{\mu} \cdot \widetilde{\lambda}) \prod_{\mu'} \delta^2(C^{\perp}_{\mu'}\cdot\lambda) \bigwedge_{\mu} (C_{\mu} \cdot d\widetilde{\lambda})^2 \bigwedge_{\mu'}(C^{\perp}_{\mu'}\cdot d\lambda)^2\,, \label{eq:form-as-contour-integral}
\end{align}
where $\mu'=3,\dots,k$, $\tilde{\mu} = 1,\dots,n{-}k$. The bosonic $\delta$-functions define a map with which allows us to pushforward the canonical form $\omega_{n,k}^\gamma$ on the Grassmannian space, 
\begin{equation}
    \omega_{n,k}^{\gamma} = F(x_1,x_2,{\dots},x_{2n-4})\,dx_1{\dots}dx_{2n-4} =\bigwedge_{i=1}^{2n-4}\frac{d\alpha_i}{\alpha_i}\,,
\end{equation}
%
to the differential form $\Omega_{n,k}^{\gamma}$ on the kinematical space. Concretely, we write $\omega_{n,k}^\gamma$ in arbitrary coordinates $\{x_i\}$ and solve $C(x)\cdot \widetilde{\lambda} = C^{\perp}(x)\cdot \lambda=0 $ for $x_i$ in terms of $\lambda,\widetilde{\lambda}$, make the substitution in the rational function $F(x_1,{\dots},x_m)$ with the appropriate Jacobian. In a canonical parameterization, eg. edge variables $\{\alpha_i\}$, the canonical form $\omega_{n,k}^{\gamma}$ is a trivial dlog and the resulting kinematical form is simply
\begin{equation}
    \Omega_{n,k}^{\gamma}(\lambda,\widetilde{\lambda}) =  \bigwedge_{i=1}^{2n-4}\frac{d\alpha_i (\lambda,\widetilde{\lambda})}{\alpha_i (\lambda,\widetilde{\lambda})}\,,
\end{equation}
where the differential operator acts on $\lambda$ and $\widetilde{\lambda}$. 

Recovering the super-function from the kinematical form $\Omega_{n,k}^\gamma$ is straightforward. Note the full super-function corresponds to an (vanishing) invariant $2n$ form. In order to recover the superfunction, we need to multiply the $2n{-}4$ form by $(dq)^4$
\begin{equation}
    {\cal F}^{\gamma}_{n,k} = (dq)^4 \wedge \Omega_{n,k}^\gamma\bigg|_{d\lambda \to \eta, d\widetilde\lambda \to \widetilde \eta}\,. \label{formF}
\end{equation}
making the replacements $d\lambda^{1,2}\rightarrow \eta^{1,2}$, $d\widetilde{\lambda}^{1,2}\rightarrow \widetilde{\eta}^{1,2}$ then turns the form back to the non-chiral superfunction ${\cal F}^{\gamma}_{n,k}$. See \cite{He:2018okq} for more details. 


The punchline: the $2n{-}4$ form $\Omega_{n,k}^\gamma$ in the kinematic space contains the same information as the super-function ${\cal F}^{\gamma}_{n,k}$ (in a non-chiral) for a given amplitude, on-shell diagram or particular BCFW cell. They can be cast into a simple dlog form that contains all this information. The translation from the form $\Omega_{n,k}^\gamma$ to a super-function is trivial, the construction of the kinematical form, however, involves non-trivial work. Let us look at a few examples.

\paragraph{MHV}

The form for the $n$-pt MHV amplitude can be famously obtained from a tringulation of the polygon, which results in 
\begin{equation}
 \input{figures/polygon_triangulation}\quad \Omega_{n,2} = \bigwedge_{i=2}^{n{-}1} d{\rm log}\frac{\la 1\,i\ra}{\la i\,i{+}1\ra}\wedge d{\rm log}\frac{\la 1\,i{+}1\ra}{\la i\,i{+}1\ra}\,,
\end{equation}
and does not depend on a particular triangulation. As noted earlier, $\Omega_{n,2}$ contains $(d\tilde{q})^4$ which is not manifest in the dlog representation. For example, at 4-point we have
\begin{equation}
    \Omega_{4,2} = \frac{(d\tilde{q})^4}{st} = d{\rm log}\frac{\la12\ra}{\la13\ra}\wedge d{\rm log}\frac{\la23\ra}{\la13\ra}\wedge d{\rm log}\frac{\la34\ra}{\la13\ra}\wedge d{\rm log}\frac{\la41\ra}{\la13\ra}\,.
\end{equation}

\paragraph{NMHV}
Consider the 6-pt NMHV cell with $(123)=0$ discussed earlier (\ref{fig:Example6pt}). Plugging the solution of C-matrix (\ref{NMHV6pt_example}) into the Grassmannian integral \eqref{eq:form-as-contour-integral} gives 
%
%
%
\begin{equation}
    \Omega_{6,3} = \frac{(d\widetilde{q})^4(d\lambda_1\la23\ra + d\lambda_2\la31\ra + d\lambda_3 \la12\ra)^2( d\widetilde{\lambda}_4[56]+d\widetilde{\lambda}_5[64] + d\widetilde{\lambda}_6[45] )^2}{s_{123}\la12\ra\la23\ra[45][56]\la1|5{+}6|4]\la3|4{+}5|6]}\,.
\end{equation}
This is equivalent to the following dlog representation
\begin{equation}
    \Omega_{6,3} = d{\rm log}(\alpha_1)\wedge d{\rm log}(\alpha_2)\wedge\dots\wedge d{\rm log}(\alpha_8)\,,
\end{equation}
with the canonical variables given by
\begin{equation}
\alpha_1 {=} \frac{\la12\ra}{\la31\ra},\,\alpha_2 {=} \frac{\la23\ra}{\la31\ra},\, \alpha_3 {=} \frac{[\hat{3}4]}{[\hat{3}\hat{1}]},\,\alpha_4 {=} \frac{[46]}{[\hat{3}\hat{1}]},\, \alpha_5{=} \frac{[6\hat{1}]}{[\hat{3}\hat{1}]},\,\alpha_6 {=} \frac{[\hat{1}4]}{[\hat{3}\hat{1}]},\,\alpha_7{=}\frac{[54]}{[64]},\,\alpha_8{=}\frac{[65]}{[64]}\,,
\end{equation}
where the shifted momenta defined as 
\begin{equation}
    \widetilde{\lambda}_{\hat{1}} = \widetilde{\lambda}_1 +  \frac{\la23\ra}{\la13\ra}\widetilde{\lambda}_2,\quad  \widetilde{\lambda}_{\hat{3}} = \widetilde{\lambda}_3 + \frac{\la12\ra}{\la13\ra}\widetilde{\lambda}_2. \label{IS_shifted_spinor}
\end{equation}
Note the dlog form then naturally splits into two parts: an anti-holomorphic dlog form for a 5-point NMHV tree-amplitude with shifted momenta, and a holomorphic dlog form for a 3-point MHV amplitude
\begin{equation}
    \Omega_{6,3} = \Omega_{5,3}(\hat{1},\hat{3},4,5,6)\wedge \Omega_{3,2}(1,2,3) .
\end{equation}
This is reminiscent of a holomorphic($k$-preserving) inverse-soft factor and indeed (\ref{IS_shifted_spinor}) is precisely the shift induced by adding a point $2$ by to the 5-pt on-shell diagram using a holomorphic inverse-soft factor. 
%
%
In general, higher point NMHV forms can be built successively from repeatedly taking holomorphic ($k$-preserving) inverse-soft factors. In particular, we can get the following dlog form representation of the general $R$-invariant ${\cal R}_{1,i{+}1,j{+}1}$ 
\begin{equation}
    \Omega_{1,i{+}1,j{+}1} = \Omega_{k=3}(\widehat{1},\widehat{i},\widehat{i{+}1},\widehat{j},\widehat{j{+}1})\wedge\Omega_{k=2}(1,{\dots},i)\wedge\Omega_{k=2}(i{+}1,{\dots},j)\wedge\Omega_{k=2}(j{+}1,{\dots},n,1) \label{SongNMHV}
\end{equation}
with the shifted $\widetilde{\lambda}$ spinors given by 
\begin{align}
 \widetilde{\lambda}_{\hat{1}} &= \widetilde{\lambda}_1 + \sum_{a=2}^{i{-}1}\frac{\la i\,a\ra}{\la i\,1\ra}\widetilde{\lambda}_a + \sum_{a=j{+}2}^n\frac{\la a\,j{+}1\ra}{\la 1\,j{+}1\ra}\widetilde{\lambda}_a, \nonumber\\
    \tilde\lambda_{\hat i} &= \tilde\lambda_i + \sum_{a=2}^{i-1}  \frac{\langle 1a\rangle}{\langle 1i \rangle}\tilde\lambda_{a}, 
   \quad
   \tilde\lambda_{\widehat{i+1}}  = \tilde\lambda_{i+1} + \sum_{a=i+2}^{j-1} \frac{\langle a, j\rangle}{\langle i+1,j\rangle}\tilde\lambda_a, 
   \nonumber
    \\
   \tilde \lambda_{\hat j} &=\tilde\lambda_{j} + \sum_{a=i+2}^{j-1}\frac{\langle i+1,a\rangle}{\langle i+1,j\rangle}\tilde\lambda_{a},  
   \quad
      \tilde\lambda_{ \widehat{j+1}} = \tilde\lambda_{j+1} + \sum_{a=j+2}^{n}\frac{\langle a,1\rangle}{\langle j+1,1\rangle}\tilde\lambda_{a}\,. 
      \label{spinor-shifts}
\end{align}
The structure of the form (\ref{SongNMHV}) reflects the way it is constructed. We start with the skeleton form $\Omega(1,i,i+1,j,j+1)$ associated with five points on three lines, 
and by successive application of holomorphic inverse-soft factors insert extra points $n, n-1,  \dots, 2$ \emph{in reversed order} to the three lines; the shifts in $\lambda$-spinors propagate through the neighboring points leading to (\ref{spinor-shifts}) in the end, and the 3-point dlog forms thus added nicely join into three $\Omega_{k=2}$ forms for the Parke-Taylor factors. For more detailed discussion on the inverse-soft construction of kinematic dlog forms see \cite{He:2018okq}. Note, however, even though the Grassmannian configuration is obvious, there is no direct geometric interpretation of (\ref{SongNMHV}) analogous to what we see in the last section. This is to be expected from the presence of the outstanding \emph{anti-holomorphic} dlog factor  $\Omega_{k=3}(\widehat{1},\widehat{i},\widehat{i{+}1},\widehat{j},\widehat{j{+}1})$.  

Differential forms for higher $k$ can also be built using the inverse-soft construction with anti-holomorphic ($k$-increasing) inverse-soft factors.  
But there is a limitation to this method. It only works for on-shell diagrams that can be built from simpler diagrams, whose form are already known, by adding points through inverse-soft factors. This is not generally the case for on-shell diagrams we encountered in the earlier discussion of BCFW cells even with adjacent shifts. Nevertheless, there is a particular recursion scheme $\{-2,2,0\}$ that yield any tree amplitude in inverse-soft constructible terms solely.

\subsection{Holomorphic dlog forms}

Now we turn our attention to the canonical dlog forms. The standard procedure is to use the Grassmannian formula (\ref{eq:form-as-contour-integral}) for a particular cell represented by a $C$-matrix. Our goal is to construct the dlog form for the general NMHV $R$-invariant ${\cal R}_{1,i{+}1,j{+}1}$ in a different way which will also generalize to the non-planar case. Inspired by the existence of the geometrical formula (\ref{R2}), we propose a \emph{holomorphic dlog form} which only depends on the holomorphic $\lambda$-spinors. The $\tilde\lambda$-dependency will be absorbed into the dependence on the spinors $\lambda_{I_1}$ and $\lambda_{I_2}$, which we defined in \eqref{eq:extra-spinors} up to a normalization. Since (\ref{R2}) gives the superfunction as the product of three Parke-Taylor factors, we may naively try to take the wedge product of three MHV dlog forms on each of the lines. It is easy to see that this cannot be exactly correct as we would get dlog forms of degree $2n-2$ instead of $2n-4$. The second problem is that the spinors $\lambda_{I_1}$ and $\lambda_{I_2}$ defined in (\ref{spinors}) were not normalized. 

The correct prescription for the holomorphic dlog form is the following. For a general NMHV configuration 
\begin{equation}
\label{ThreeLines2a}
\input{figures/ThreeLines2}
\end{equation}
take the wedge product of three dlog forms
\begin{equation}
    \Omega_{n,k=3} = \widetilde{\Omega}_{k=2}(1,2,\dots,i,I_1)\wedge\Omega_{k=2}(I_1,i{+}1,{\dots},j,I_2)\wedge\widetilde{\Omega}_{k=2}(I_2,j{+}1,{\dots},n,1).
\end{equation}
The spinors for $I_1$ and $I_2$ are defined with correct normalization as follows:
\begin{equation}
    \lambda_{I_1} = \la 1|P_3P_2 \cdot \frac{\la 1i\ra}{\la 1|P_1P_3|1\ra},\qquad \lambda_{I_2} = \la 1|P_1P_2\cdot \frac{\la 1\,j{+}1\ra}{\la 1|P_1P_3|1\ra}\,. \label{spinors2}
\end{equation}

The dlog form for the middle line is the usual MHV amplitude for points $I_2,i{+}1,{\dots},j,I_1$ and it is given by the triangulation of a corresponding polygon
\begin{align}
\input{figures/Triang1}
\qquad 
\begin{aligned}
        &\Omega_{k=2}(I_1,i{+}1,{\dots},j,I_2)   \\ 
        = \ &\Omega_{k=2}(i{+}1,{\dots},j)\wedge d{\rm log}\frac{\la i{+}1\,j\ra}{\la jI_2\ra}\wedge d{\rm log}\frac{\la i{+}1\,I_2\ra}{\la jI_2\ra} \\
    & \hspace{2.6 cm}  \wedge d{\rm log}\frac{\la i{+}1\,I_2\ra}{\la I_2I_1\ra}\wedge d{\rm log}\frac{\la i{+}1\,I_1\ra}{\la I_2I_1\ra}\,,
\end{aligned}
\label{omega1}
\end{align}
where we denoted the dlog form for the polygon $(i{+}1,\dots,j)$,
\begin{equation}
   \Omega_{k=2}(i{+}1,{\dots},j) = \bigwedge_{k=i{+}2}^{j{-}1} d{\rm log}\frac{\la i{+}1\,k\ra}{\la k\,k{+}1\ra}\wedge d{\rm log}\frac{\la i{+}1\,k{+}1\ra}{\la k\,k{+}1\ra}\,,
\end{equation}
and we added in (\ref{omega1}) dlog forms for two triangles $(i{+}1,j,I_2)$ and $(i{+}1,I_2,I_1)$ which depend on extra points $I_1$, $I_2$. This all trivially follows from the triangulation. 

The \emph{reduced} dlog form $\widetilde{\Omega}_{k=2}(1,2,\dots,i,I_1)$ also corresponds to a polygon,
\input{figures/Triang2}
We triangulate the sub-polygon with points $(1,2,{\dots},i)$ in the standard way and write its dlog form. For the last triangle $(1,i,I_1)$ we only associate one dlog factor rather than two,
\begin{equation}
    \widetilde{\Omega}_{k=2}(1,2,\dots,i,I_1) = \Omega_{k=2}(1,2,{\dots},i)\wedge d{\rm log}\frac{\la I_1\,i\ra}{\la I_1\,1\ra}\,.
\end{equation}
Interestingly the last triangle is very degenerate, therefore there is indeed only one non-trivial ratio we can construct as $\la 1I_1\ra = \la 1i\ra$. In some sense it is a dlog form on the line rather than a ``half-dlog form" on a triangle. Similarly, the other reduced dlog form is 
\begin{equation}
    \widetilde{\Omega}_{k=2}(I_2,j{+}1,{\dots},n,1) = \Omega_{k=2}(j{+}1,\dots,n,1)\wedge d{\rm log}\frac{\la I_2\,j{+}1\ra}{\la I_2\,1\ra}
\end{equation}
Everything combined, we can write the holomorphic dlog form for the general NMHV configuration associated with the $R$-invariant ${\cal R}_{1,i{+}1,j{+}1}$ as
\begin{align}
    \Omega_{k=3} &=\Omega_{k=2}(1,{\dots},i)\wedge \Omega_{k=2}(I_1,i{+}1,{\dots},j,I_2)\wedge \Omega_{k=2}(j{+}1,{\dots},n,1) \nonumber\\
    & \quad\wedge d{\rm log}\frac{\la I_1\,i\ra}{\la I_1\,1\ra} \wedge d{\rm log}\frac{\la I_2\,j{+}1\ra}{\la I_2\,1\ra}\,,\label{NMHVdlog}
\end{align}
which is the wedge product of three MHV amplitude dlog forms with two extra dlogs. We checked explicitly that the formula (\ref{NMHVdlog}) is equivalent to (\ref{SongNMHV}) .

As an example, the 8-dimensional cell in $G_+(3,6)$ for $(123)=0$ turns into 
\begin{equation}
    \Omega_{6,3} = \Omega_{k=2}(1,2,3)\wedge \Omega_{k=2}(I_1,4,5,I_2)\wedge d{\rm log}\frac{\la13\ra}{\la 3I_1\ra}\wedge d{\rm log}\frac{\la 16\ra}{\la 1I_2\ra}\,,
\end{equation}
where the $I_1$ and $I_2$ $\lambda$-spinors are defined as
\begin{equation}
    \lambda_{I_1} = (45)|6]\cdot\frac{\la 13\ra}{\la 1|23|6]},\qquad \lambda_{I_2} = \la 1|(23)(45)\cdot \frac{\la 16\ra}{\la 1|(23)(45)|1\ra}\,.
\end{equation}

Our construction uses only holomorphic data and the dlog form is a wedge product of rational functions which only depend on the $\lambda$-part of $1,2,\dots,n$ and $I_1,I_2$. Obviously, the final form must also have a dependence on $\widetilde{\lambda}$s of external momenta, which comes exclusively through the dependence on $\lambda_{I_1}$ and $\lambda_{I_2}$. While the forms in \cite{He:2018okq} and the ones presented here, are completely equivalent, the holomorphicity of our representation is indicative of the geometric fact that the answer was built from the dlog forms on three lines. This underscores the notion that we can think about the Grassmannian geometry picture directly in the kinematical space. It is suggestive that the same construction can be used for higher $k$ to build the N$^k$MHV dlog form from the MHV dlog forms on individual lines.

\subsection{Non-planar generalization}

The formula for the dlog form from the previous section directly generalizes to the non-adjacent BCFW terms and the non-planar positive geometry. Let us directly consider the general case,  
%
%

\input{./figures/DoubleLine.tex}

The strategy is the same as before, write the dlog form for the middle line and partial dlog forms on the other two lines with two compensating terms. On the first line, the points are cyclically ordered $(1,2,{\dots},i,I_1,i{+}1,{\dots},j) = (i{+}1,{\dots},j,1,2,{\dots},i,I_1)$. The final formula for the dlog form is 
\begin{align}
    \Omega &=\Omega_{k{=}2}(i{+}1,{\dots},j,1,2,{\dots},i)\wedge \Omega_{k{=}2}(I_1,j{+}1,{\dots},k,I_2,k{+}1,{\dots},m)\nonumber\\
    & \hspace{2cm}\wedge \Omega_{k{=}2}(m{+}1,{\dots},n)\wedge d{\rm log}\frac{\la I_1\,i\ra}{\la I_1\,i{+}1\ra}\wedge d{\rm log}\frac{\la I_2\,m{+}1\ra}{\la I_2\,1\ra}\,, \label{dlog2}
\end{align}
where the spinors $\lambda_{I_1}$ and $\lambda_{I_2}$ are given by (\ref{spinors2}),
\begin{equation}
    \lambda_{I_1} = \la 1|P_3P_2\cdot \frac{\la 1\,i\ra}{\la 1|P_1P_3|1\ra},\qquad \lambda_{I_2} = \la 1|P_1P_2 \cdot \frac{\la 1\,j{+}1\ra}{\la 1|P_1P_3|1\ra}\,,
\end{equation}
with $P_1$, $P_2$, $P_3$ being the sums of momenta on the three lines excluding $p_1$. The formula for the dlog form (\ref{dlog2}) is basically identical to the planar counterpart (\ref{NMHVdlog}), though we had to be careful about which extra dlog factor to add to the first line. Note that the argument here is $\la I_1\,i\ra/\la I_1\,i{+}1\ra$ from the triangle $(i,I_1,i{+}1)$. This triangle is not degenerate, so we can not use $\la i\,i{+}1\ra$ in the argument (this is indeed not a pole). 

As an example, we do two 6-point non-planar configurations,
\input{./figures/Examples3.tex}
In the first case, we get
\begin{equation}
    \Omega = \Omega_{k{=}2}(1,2,3)\wedge \Omega_{k{=}2}(I_1,4,I_2,5)\wedge d{\rm log}\frac{\la I_1\,3\ra}{\la I_1\,1\ra}\wedge d{\rm log}\frac{\la I_2\,6\ra}{\la I_2\,1\ra}\,,
\end{equation}
where $I_1$ and $I_2$ spinors are
\begin{equation}
    \lambda_{I_1} = (45)|6]\cdot \frac{\la 13\ra}{\la 1|23|6]},\qquad \lambda_{I_2} = \la 1|(23)(45) \cdot \frac{1}{\la 1|23|6]}\,.
\end{equation}
In the second example, the dlog form is 
\begin{equation}
    \Omega = \Omega_{k{=}2}(I_1,3,4,I_2,5)\wedge d{\rm log}\frac{\la I_1\,2\ra}{\la I_1\,1\ra}\wedge d{\rm log}\frac{\la I_2\,6\ra}{\la I_2\,1\ra}\,,
\end{equation}
where
\begin{equation}
    \lambda_{I_1} = (12)|6]\cdot \frac{1}{[26]},\qquad \lambda_{I_2} = (16)|2] \cdot \frac{1}{[26]}\,.
\end{equation}
As a result, we conclude that the same formula for the holomorphic dlog form works for the planar (\ref{NMHVdlog}) as well as the non-planar (\ref{dlog2}) cases. The planarity/convexity of the configuration does not matter; the expression only depends on the Grassmannian configuration of points on three lines in $\mathbb{P}^2$. The holomorphicity of the dlog form (\ref{dlog2}) and the representation of the superfunction using Parke-Taylor factors (\ref{Rinv4}) shows that we can indeed think about the Grassmannian configuration of points in $\mathbb{P}^2$ as the collection of three lines directly in the kinematical $\lambda$-space.

\section{N\texorpdfstring{$^2$}{\texttwosuperior}MHV and Beyond}
\label{sec:higherk}
In the previous sections, we restricted our discussion to MHV and NMHV cases. In this section, we generalize our discussion of Grassmannian geometry for BCFW cells to N$^2$MHV, considering both adjacent and non-adjacent shifts. The same construction extends to higher $k$ and we outline how it works for arbitrary $k$.

\subsection{N\texorpdfstring{$^2$}{\texttwosuperior}MHV geometries}
\subsubsection*{Planar geometries}
Let us first consider the adjacent BCFW shift $(n1)$. At N$^2$MHV, we have three types of terms in the recursion,
\begin{equation}
\input{./figures/N2MHVterms.tex} \label{N2MHVterms}
\end{equation}
As before, we can express these terms as sums of on-shell diagrams. For each diagram we construct the representative $C$-matrix and calculate the superfunction as a Grassmannian integral. Now the $C$-matrix, viewed as a collection of $n$ columns, describes a configuration of $n$ points in $\mathbb{P}^3$. The Grassmannian geometries associated with factorizations into  MHV and NMHV amplitudes can be built from the MHV and NMHV configurations we found in the previous sections. In the first term in (\ref{N2MHVterms}), we have a MHV amplitude $A^{k=2}(I,i{+}1,{\dots},n{-}1,\hat{n})$, represented by a line, and an NMHV amplitude $A^{k=3}(I,\hat{1},2,{\dots},i)$, given by a sum over configurations of points localized on three lines (where we sum over all $j,k$), 
\begin{equation}
\input{./figures/N2MHVglue1.tex}   \label{N2MHVglue1}
\end{equation}
Gluing the MHV and NMHV configurations together at points $1$ and $I$ leads to N$^2$MHV configurations of $n$ points on five lines in $\mathbb{P}^3$. Merging the MHV line $A^{k=2}(1,2,{\dots},i,I)$ with the collection of NMHV configurations $A^{k=3}(I,i{+}1,{\dots},n{-}1,\hat{n})$ gives the second term in (\ref{N2MHVterms}). As a result, we get a different configuration of $n$ points on five lines in $\mathbb{P}^3$, 
\begin{equation}
    \input{./figures/N2MHVglue2.tex}  \label{N2MHVglue2}
\end{equation}
Moreover, from the recursive argument geometries for the last term in (\ref{N2MHVterms}) are obtained by adding more points on the $\hat{n},n,1$ line of the N$^2$MHV configuration. This is similar to the NMHV case. The resulting configurations complement those associated with the first two terms and in the end, the N$^2$MHV amplitude can be written as a sum over on-shell functions corresponding to the cells in $G_+(4,n)$ represented by two types of configurations,
\begin{equation}
    \input{./figures/N2MHVgeneral_1.tex} \qquad \input{./figures/N2MHVgeneral_2.tex} \label{N2MHVgeneral}
\end{equation}
where we sum over all $1<j<k<i<m<n$ and $1<i<j<k<m<n$ labels respectively (in the ordering indicated in the figures, subject to the constraint that there are always at least two points on each line). All these $(2n{-}4)$-dimensional configurations are \emph{convex}, ie. all $(4\times4)$ ordered minors of the $C$ matrix are positive (or zero). 
\subsubsection*{Non-planar geometries}

We can now extend our discussion of Grassmannian geometries 
for non-adjacent BCFW shifts to N$^2$MHV. The non-adjacent BCFW shifts produce three types of terms in the recursion relations, differing from (\ref{N2MHVterms}) only by having $k,1$ non-consecutive:
\input{figures/N2MHVnon1.tex}
Here the last term includes the BCFW term with $\mathcal{A}_3^{k=1}(\hat{k},I,k-1)$ on the left and the one with $\mathcal{A}_3^{k=1}(\hat{k},k+1,I)$ on the left.
Associated with the first term is a sum of configurations arising from merging NMHV and MHV configurations, 
\begin{equation}
\input{./figures/N2MHVnon2.tex}  \label{N2MHVnon2}
\end{equation}
where we only denoted labels as necessary for make the gluing procedure clear. All other labels are inherited from the MHV and NMHV configurations. Note that the only difference between (\ref{N2MHVglue1}) and (\ref{N2MHVnon2}) is that some points are ``misplaced" on the lines, ie. they are on both sides of the intersections points with other lines, which makes the configurations non-convex. Hence these are configurations in the general Grassmannian $G(4,n)$, rather than in the positive part $G_+(4,n)$. Similarly, for the second term we get
\begin{equation}
\input{./figures/N2MHVnon3.tex} \label{N2MHVnon3}
\end{equation}
Finally, the third term, which is of the form ($(n{-}1)$-point N$^2$MHV) $\times$ (3-point $\overline{\rm MHV}$) corresponds to adding more points on the line connecting points $k$ and $1$. As in the planar case, this can be shown recursively. In the end we get two types of configurations for the N$^2$MHV amplitude
\begin{equation}
\input{./figures/N2MHVnon4.tex}  \label{N2MHVnon4}
\end{equation}
The BCFW expansion instructs us to sum over both topologies and the distribution of all labels which respect the relative ordering of $2,\dots,k{-}1$ and $k{+}1,\dots,n$. The labels flows are then
\begin{equation}
\input{./figures/N2MHVnon5.tex} \label{N2MHVnon5}
\end{equation}
where the blue arrow represents points $ 2,\dots,k{-}1$ and the red arrow points $k{+}1,\dots,n$ (or vice versa). The red arrows go in order $1,2,3$. The geometry configuration is projective (unlike our drawing), so the picture is actually symmetric in switching red and blue arrows (finite $\leftrightarrow$ infinite intervals). Note that on three of the lines, the red and blue labels do not overlap, while on the remaining two lines they are mixed together, respecting the shuffle product ordering.

For higher $k$, we use the same procedure to glue together two geometries into a higher-dimensional configuration. For example, at N$^3$MHV the BCFW recursion for a $(k1)$ shift contains a following term,
\begin{equation}
\input{figures/N3MHVnon1.tex} \label{N3MHVnon1}
\end{equation}
This gives rise to a configuration of $n$ points in $\mathbb{P}^4$,
\begin{equation}
\input{figures/N3MHVnon2.tex} \label{N3MHVnon2}
\end{equation}
This is hard to visualize, but we can see here two projective planes $\mathbb{P}^2$, which are glued together with an additional line $(1k\hat{k})$ that lives in an additional direction, making the space $\mathbb{P}^4$ rather than $\mathbb{P}^3$. We can see that four of the lines have misplaced points, this generalizes to higher $k$ as well. For general $k$, we have configurations of $n$ points on $2k-3$ lines in $\mathbb{P}^{k-1}$ where $k{-}1$ lines have misplaced points.

\subsection{Holomorphic on-shell functions}
From our study of MHV and NMHV amplitudes, we have seen that Grassmannian configurations, with the kinematic space interpretation, provides a fast track to the on-shell function. 
The geometrical formula expresses an NMHV amplitudes as a product of Parke-Taylor factors with auxiliary labels which encode special points in the Grassmannian configurations. Our goal in this subsection is to generalize this formula to N$^2$ MHV and beyond.
\subsubsection*{On-shell functions from planar configurations}

Let us begin with the planar configurations (\ref{N2MHVgeneral}). Note that the five lines on which the points are localized lie in two different planes and each plane has three lines (with one line overlapping). This is precisely the structure encoded in the representation of the N$^2$MHV on-shell functions as a product of two $R$-invariants, \cite{Drummond:2008cr}, 
\begin{equation}
    {\cal F} = \frac{\delta^4(P)\delta^8(Q)}{\la12\ra\la23\ra\dots \la n1\ra}\times R[a_1,a_2,a_3,a_4,a_5]\,R[b_1,b_2,b_3,b_4,b_5]\,, \label{eq:N2MHV-RR}
\end{equation}
with each R-invariant being associated with on one of the planes. The arguments of the $R$-invariants are shifted (super-)momentum twistors; the same entries can also be thought of as non-zero entries on the rows in the positive Grassmannian $G_+(4,n)$. This formula generalizes to higher $k$ where each $(2n{-}4)$-dimensional configuration of $n$ points in $\mathbb{P}^{k{-}1}$ (localized on $2k-3$ lines) gives rise to an on-shell function which can be expressed as a product of $k$ copies of $R$-invariants with various shifted indices. Together they furnish a representation of the N$^{k-2}$MHV amplitude in terms of $R$-invariant products. For more details of this representation see \cite{Drummond:2008cr}. 

Now back to the kinematic space of $\{\lambda, \tilde\lambda,  \eta\}$. The formula \eqref{eq:N2MHV-RR} can be rewritten in terms of $\lambda$-spinors only with the introduction of some auxiliary spinors which correspond to special points in the Grassmannian configurations (\ref{N2MHVgeneral}). The holomorphic expression consists of five Parke-Taylor factors.
We start with the first convex configuration, and add intersection points $I_1$, $I_2$ and $I_3$ in the figure,
\begin{equation}
\input{./figures/N2MHVgeneral1.tex} \label{N2MHVgeneral1}
\end{equation}
To set up notation, we collectively label momenta on individual lines,
\begin{equation}
    P_1^a = p_2{+}{\dots}{+}p_j,\quad P_1^b = p_{j{+}1}{+}{\dots}{+}p_k,\quad P_1^c = p_{k{+}1}{+}{\dots}{+}p_i\,,
\end{equation}
\begin{equation}
    P_2 = p_{i{+}1}{+}{\dots}{+}p_m,\quad P_3 = p_{m{+}1}{+}{\dots}{+}p_n\,.
\end{equation}
We also define $P_1=P_1^a+P_1^b+P_1^c$. We need to associate momenta to the points $I$, $I_1$, $I_2$ and $I_3$. The momentum $p_I$ can be easily obtained directly from the BCFW term,
\begin{equation}
 p_I = \frac{\la 1|P_3P_2 \times\la 1|P_1}{\la 1|P_1P_3|1\ra}\,.
\end{equation}
All other $\lambda$-spinors can be directly read off from the geometry following the same rules as we identified for $I_1$ and $I_2$ in the NMHV case. The first plane gives us,
\begin{equation}
    \lambda_{I_2} = \la 1|P_1^aP_1^b,\qquad \lambda_{I_1} = \la 1|(P_1^c+p_I)P_1^b = \la 1|(P_1^a+P_1^b)P_1^b\,.
\end{equation}
Similarly we can calculate $\lambda_{I_3}$, making the list of all $\lambda$-spinors we need 
\begin{equation}
    \lambda_I = \la1|P_3P_2,\quad \lambda_{I_1} =\la1|(P_1^a+P_1^b)P_1^b,\quad \lambda_{I_2} = \la 1|P_1^aP_1^b,\quad \lambda_{I_3} = \la 1|P_1P_2\,.
\end{equation}
The superfunction which can be computed from (\ref{dual}) takes the familiar form of a product of Parke-Taylor factors,
\begin{align}
    {\cal F} &= PT(1,2,{\dots},j,I_1)\times PT(I_1,j{+}1,{\dots},k,I_2)\times PT(I_2,k{+}1,{\dots},i,I,1)\nonumber\\
    &\hspace{0.3cm}\times PT(I,i{+}1,{\dots},m,I_3)\times PT(I_3,m{+}1,{\dots},n,1) \nonumber\\
    & \hspace{0.3cm}\times \la 1|P_1P_2\ra^3\cdot \la 1|P_1^aP_1^b|1\ra^3\cdot \delta^8(\Xi)\,.
\end{align}
Note that we have two normalization factors coming from each of the planes $\la 1|P_1P_2\ra^3$ and $\la 1|P_1^aP_1^b|1\ra^3$. Also, $\la1I\ra=\la 1I_3\ra = \la1|P_2P_3|1\ra$ and $\la 1I_1\ra=\la1I_2\ra = \la 1|P_1^aP_1^b|1\ra$. We have factored out and omitted momentum and super momentum-conserving delta functions $\delta^4(P)\delta^8(Q)$ as usual.
%
This can always be achieved by setting the first two rows of the $C$-matrix to $\lambda$ with $GL(4)$ gauge symmetry. What remains of the fermionic constraints $\delta^{4\times 4}(C\cdot\eta)$ evaluates to $\delta^8(\Xi)$, which is a product of two NMHV delta functions,
\begin{equation}
    \delta^8(\Xi) = \delta^4(\Xi_1)\,\delta^4(\Xi_2)\,.
\end{equation}
corresponding to the third and fourth rows of the localized $C$-matrix. As one may expect, the expressions for $\Xi_1$ and $\Xi_2$ can be directly read from the Grassmannian geometry, focusing on one plane at a time,
\begin{equation}
    \Xi_1 = \sum_{r\in P_1^b}\la r|P_1^bP_1^a|1\ra\eta_r + \sum_{s\in P_1^a} \la 1s\ra(P_1^b)^2\eta_s,\qquad
    \Xi_2 = \sum_{r\in P_2}\la r|P_2P_3|1\ra\eta_r + \sum_{s\in P_3}\la 1s\ra P_2^2\eta_s\,. \label{Xi1}
\end{equation}
Note that there are many ways to write the delta function on the support of super momentum conservation. As before, the arguments $\Xi_1$ and $\Xi_2$ do not depend on a particular ordering of points on the given lines.

The second configuration can be analyzed in a similar way
\begin{equation}
\input{./figures/N2MHVgeneral2.tex} 
\end{equation}
Denote
\begin{align}
    P_1 &= p_2{+}{\dots}{+}p_i,\quad P_2^a = p_{i{+}1}{+}{\dots}{+}p_j,\quad P_2^b = p{j{+}1}{+}{\dots}{+}p_k,\nonumber\\
    & \hspace{1cm}P_2^c = p_{k{+}1}{+}{\dots}{+}p_m,\quad P_3=p_{m{+}1}{+}{\dots}{+}p_n\,.
\end{align}
The momentum $p_I$ in these conventions is
\begin{equation}
    p_I = \frac{\la 1|P_3P_2\times \la 1|P_1}{\la1|P_1P_3|1\ra}\,,
\end{equation}
and $\lambda_I = \la 1|P_3P_2$. We can solve for the $\lambda_{I_3}$ spinor from the first plane, to get $\lambda_{I_3} = \la1|P_1P_2$. The other two spinors can be obtained from the second plane, but now the ``center'' is label $I$ rather than $1$,
\begin{equation}
    \lambda_{I_1} = \la I|(P_2^a+P_2^b)P_2^b,\qquad \lambda_{I_2} = \la I|P_2^aP_2^b\,.
\end{equation}
The resulting superfunction is then 
\begin{align}
    {\cal F} &= PT(1,2,{\dots},i,I)\times PT(I,i{+}1,{\dots},j,I_1)\times PT(I_1,j{+}1,{\dots},k,I_2)\nonumber\\
    &\hspace{0.2cm}\times PT(I_2,k{+}1,{\dots},m,I_3,I)\times PT(I_3,m{+}1,{\dots},n,1) \nonumber \\ 
    & \hspace{0.2cm} \times \la 1|P_1P_3|1\ra^3\cdot \la I|P_2^aP_2^b|I\ra^3\cdot \delta^8(\Xi)\,. \label{N2MHVfun0}
\end{align}
Following the same logic, we can factorize the fermionic delta function $\delta^8(\Xi){=}\delta^4(\Xi_1)\delta^4(\Xi_2)$, where
\begin{equation}
    \Xi_1 = \sum_{r\in P_3}\la r|P_3P_1|I\ra\eta_r + \sum_{s\in P_1}\la Is\ra P_3^2\eta_s,\qquad
    \Xi_2 = \sum_{r\in P_2^b}\la r|P_2^bP_2^a|I\ra\eta_r + \sum_{s\in P_2^a} \la Is\ra(P_2^b)^2\eta_s\,.
\end{equation}
Note that we used the supermomentum conservation in $\Xi_1$ to eliminate $\eta_{I_2}$. 

\subsubsection*{On-shell functions from non-planar configurations}

The configurations stemming from non-adjacent BCFW shifts follow an identical pattern: associate momenta for all intersection points $I,I_1,I_2,I_3$ and take the product of Parke-Taylor factors, one for each line. 
\begin{equation}
    \input{./figures/General4.tex} \label{N2MHVgen} 
\end{equation}
\begin{align}
    {\cal F} &= PT(1,2,{\dots},i,I_1,i{+}1,{\dots},j)\times PT(I_1,j{+}1,{\dots},k,I_2,k{+}1,{\dots},l)\nonumber\\
    &\hspace{0.2cm}\times PT(I_2,l{+}1,{\dots},m,I,1)\times PT(I,m{+}1,{\dots},p,I_3,p{+}1,{\dots},q)\times PT(I_3,q{+}1,{\dots},n,1)\nonumber\\
    &\hspace{0.2cm}\times \la 1|P_1P_2\ra^3\cdot \la 1|P_1^aP_1^b|1\ra^3\cdot \delta^8(\Xi)\,,\label{N2MHVfun}
\end{align}
where the definitions of the variables $P_1^a$, $P_1^b$, $P_1^c$, $P_1$, $P_2$, $P_3$, $I$, $I_1$, $I_2$, $I_3$, $\Xi$ are the same as in the planar configuration. Hence the normalization factors and the fermionic $\delta$-functions are identical. The only difference is the ordering of points in the Parke-Taylor factors. 

The on-shell function (\ref{N2MHVfun}) is a generalization of N$^2$MHV Yangian invariants. Similar to the NMHV case, we can now decompose (\ref{N2MHVfun}) into a linear combination of (\ref{N2MHVfun0}) with different orderings by rewriting the Parke-Taylor factors in (\ref{N2MHVfun}). In (\ref{N2MHVgen}), points are now misplaced on three of the lines. We use KK relations to rewrite the corresponding Parke-Taylor factors using terms which have ($1$, $I_1$), ($I_1$, $I_2$) and ($I$, $I_3$) adjacent,
\begin{align}
   PT(1,2,{\dots},i,I_1,i{+}1,{\dots},j) &= \sum_{\sigma_1\in \Sigma_1} PT(1,\sigma_1,I_1)\,,\\
   PT(I_1,j{+}1,{\dots},k,I_2,k{+}1,\dots,l) &= \sum_{\sigma_2\in \Sigma_2} PT(I_1,\sigma_2,I_2)\,,\\
   PT(I,m{+}1,{\dots},p,I_2,p{+}1,\dots,q) &= \sum_{\sigma_3\in \Sigma_3} PT(I,\sigma_3,I_3)\,,
\end{align}
where $\Sigma_1=\{1,2,{\dots},i\}\,\shuffle\,\{i{+}1,{\dots},j\}^T$, $\Sigma_2=\{j{+}1,{\dots},k\}\,\shuffle\,\{k{+}1,\dots,l\}^T$ and finally $\Sigma_3=\{m{+}1,{\dots},p\}\,\shuffle\,\{p{+}1,\dots,q\}^T$. As before, $\shuffle$ denotes a shuffle product and $\{\}^T$ denotes reversal of ordering.
Plugging back into (\ref{N2MHVfun}) we get a linear combination of Yangian-invariants, 
\begin{equation}
    {\cal F} = \sum_{\sigma_1,\sigma_2,\sigma_3} {\cal F}(1,\sigma_1,\sigma_2,l{+}1,{\dots},m,\sigma_3,q{+}1,\dots,n) = \input{figures/General5} \label{General5}
\end{equation}
%
%

\paragraph{Higher $\mathbf{k}$}
Our procedure for on-shell functions extends to higher $k$. We conjecture that regardless of planarity, an on-shell function that originates from BCFW recursion has the following structure
\begin{equation}
    {\cal F} = \prod_{\rm lines} PT(\dots) \times \prod_{\rm planes} J \times \prod_{\rm planes} \delta^4(\Xi)\,.
\end{equation}
For any non-planar N$^k$MHV configuration that originates from non-adjacent BCFW shifts, applying KK relations to its Parke-Taylor factors allows us to rewrite the on-shell function as a linear combination of Yangian-invariants, each of which correspond to planar configurations with various orderings.

\section{Conclusions and Outlook}
\label{sec:discuss}

In this paper, we discussed non-adjacent BCFW recursion relations and the connection between individual terms and Grassmannian geometry. We found that each term corresponds to a particular cell in the Grassmannian $G(k,n)$ which we identified using configurations of $n$ points in $\mathbb{P}^{k{-}1}$. These Grassmannian cells are similar to the $2n{-}4$ dimensional cells in the positive Grassmannian $G_+(k,n)$ which are associated with adjacent BCFW terms, but their underlying configurations of points do not have one fixed ordering. In fact, there are two separate orderings of all points which lie between the two shifted legs $1$ and $k$. We mostly focused on the NMHV case, extending the analysis of MHV on-shell diagrams and geometries in \cite{Arkani-Hamed:2014via}. We outlined a general structure in Section \ref{sec:higherk}.

We also found a new representation of the on-shell functions associated with these geometries, both planar and non-planar, as the product of Parke-Taylor factors localized on each line in Grassmannian space. We identified momenta associated with intersection points $I_k$, which then entered the arguments of the Parke-Taylor factors. As a result, the on-shell forms were purely holomorphic in $\lambda$ and the only non-holomorphicity (dependence on $\widetilde{\lambda}$) came from the dependence on special points $\lambda_{I_k}$. This is a very interesting picture because it allows us to think about the Grassmannian geometry directly in the momentum space. This connection is trivial for the MHV case but it seems to have a lot of non-trivial structure for higher helicity sectors. The representation using Parke-Taylor factors allowed us to use Kleiss-Kuijf relations and rewrite any BCFW non-planar superfunction as a linear combination of planar super functions with various orderings. For a fixed ordering, the on-shell superfunction can be written as a product of $R$-invariants and enjoys the infinite-dimensional Yangian symmetry. Whether the Yangian structure survives in any form beyond the planar limit where the ordering is lost is an important open question, and the simple relation between the planar and non-planar on-shell functions offers a window into future investigation.

Our work opens a new direction of study of non-planar on-shell functions and geometries in ${\cal N}=4$ SYM theory. The BCFW building blocks belong to a larger family of non-planar on-shell diagrams, which appear as cuts of loop integrands to any loop order. Each non-planar on-shell diagram is associated with some Grassmannian geometry, but the concrete description is unclear. We made a first step in this direction by identifying the Grassmannian geometries associated with BCFW cells. It would be very interesting to learn what the positive geometries for all non-planar on-shell diagrams are and find a direct way to evaluate the on-shell functions. This work was partially started for the simplest $G(3,6)$ case where all on-shell diagrams corresponding to 9-dimensional and 8-dimensional cells were classified (including their canonical forms). The connection of these on-shell diagrams to Grassmannian geometries of points in $\mathbb{P}^2$ is work in progress \cite{progress}. The bigger goal of this effort is to understand how much geometric structure and what hidden symmetries and mathematical properties extend beyond the planar ${\cal N}=4$ SYM amplitudes.

\vspace{-0.3cm}

\acknowledgments

\vspace{-0.3cm}

We thank Nima Arkani-Hamed for useful discussions and comments. S.P., J.T and M.Z. are supported by a DOE grant No. SC0009999 and the funds of the University of California.

\bibliographystyle{JHEP}
\bibliography{refs.bib}

\end{document}

%% file: figures/Triang2.tex
\begin{align}
    \begin{tikzpicture}[ baseline={(0,1cm)}]
    \draw(0,0.5)--(-1,1) node[at start, below] {$1$} node[at end, left] {$2$};
    \draw[dashed] (-1,1)--(-1,2) node[at end, left] {$i-1$};
    \draw(-1,2)--(0,2.5) node[at end, above] {$i$};
    \draw[pattern=horizontal lines] (0,2.5)--(0,0.5)--(1,1.5) node[right] {$I_1$}--cycle;
    \end{tikzpicture}
\end{align}